\documentclass[aps,prd,groupedaddress,twocolumn,floatfix,showpacs,tightenlines]{revtex4-1}
\usepackage{graphicx}
\usepackage{color}
\usepackage{amsmath}

\newcommand{\beq}{\begin{equation}}
\newcommand{\eeq}{\end{equation}}
\newcommand{\bea}{\begin{eqnarray}}
\newcommand{\eea}{\end{eqnarray}}
\newcommand{\bes}{\begin{subequations}}
\newcommand{\ees}{\end{subequations}}

\newcommand{\KMS}{\rm km\ s^{-1}}
\providecommand{\e}[1]{\ensuremath{\times 10^{#1}}}

\begin{document}
\title{Kicking gravitational wave detectors with recoiling black holes}
\author{Carlos O. Lousto}
\author{James Healy} 
\affiliation{Center for Computational Relativity and Gravitation,\\
School of Mathematical Sciences,
Rochester Institute of Technology, 85 Lomb Memorial Drive, Rochester,
New York 14623}
\date{\today}

\begin{abstract}
  Binary black holes emit gravitational radiation with net linear momentum leading to a retreat of the final remnant black hole that can reach up to $\sim5,000$ km/s. Full numerical relativity simulations are the only tool to accurately compute these recoils since they are largely produced when the black hole horizons are about to merge and they are strongly dependent on their spin orientations at that moment. We present eight new numerical simulations of BBH in the hangup-kick configuration family, leading to the maximum recoil. Black holes are equal mass and near maximally spinning ($|\vec{S}_{1,2}|/m_{1,2}^2=0.97$). Depending on their phase at merger, this family leads to $\sim\pm4,700$ km/s and all intermediate values of the recoil along the orbital angular momentum of the binary system. We introduce a new invariant method to evaluate the recoil dependence on the merger phase via the waveform peak amplitude used as a reference phase angle and compare it with previous definitions.
  We also compute the mismatch between these hangup-kick waveforms to infer their observable differentiability by gravitational wave detectors, such as advanced LIGO, finding currently reachable signal-to-noise ratios, hence allowing for the identification of highly recoiling black holes having otherwise essentially the same binary parameters.
\end{abstract}

\maketitle


\section{Introduction}\label{sec:intro}

Soon after numerical relativity simulations \cite{Campanelli:2007ew,Campanelli:2007cga}
neatly revealed that astrophysical binary black holes may impart speeds of
thousands of kilometers per seconds after merger on the final black hole through gravitational recoil,
a search for them intensified in the astronomical community.
These searches ranged from
dynamical effects of their host galaxies \cite{Volonteri:2007et,Loeb:2007wz,HolleyBockelmann:2007eh,sesana:2007zk,Blecha:2008mg} leading to displacements from galaxy cores, to specific objects displaying features that could be interpreted as
differential velocities of thousand of kilometers per second between narrow and broad emission lines, like
CID-42 \cite{Civano:2010es,Blecha:2012kx,Lanzuisi:2013fza},
J0927+2943 \cite{Bogdanovic:2008uz,Komossa:2008qd,Vivek:2009mm,Shields:2008kn,Decarli:2014wca},
J1225+1415 \cite{Jonker:2010ip},
J1050+3456 \cite{Shields:2009jf}, and
NGC1277 \cite{Shields:2013yaa}. 
A systematic search was carried out and described in \cite{Eracleous:2011ua,Lena:2014zpa,Runnoe:2015yba,Runnoe:2017oxn}.
A particularly promising study of 3C186 \cite{Chiaberge:2016eqf,Lousto:2017uav,Chiaberge:2018lkg} is currently underway.
Early reviews on this field are given in Refs. \cite{Komossa:2012cy,Blecha:2015baa}.

Systematic numerical relativity simulations provided a method to model the recoil of
the final merged black hole as a function of the precursor binary \cite{Lousto:2012su,Zlochower:2015wga},
and to determine that the maximum recoil is about 5,000 km/s for maximally
spinning, equal mass, holes in the hangup kick configuration \cite{Lousto:2011kp}.
Aligned spins, on the other hand, can only reach a maximum of just above 500 km/s,
in an antialigned configuration with mass ratio $q\sim2/3$ \cite{Healy:2014yta,Healy:2017vuz}. While nonspinning holes only contribute about one third of this value \cite{Gonzalez:2006md,Healy:2017mvh}. See a review of the field in \cite{Sperhake:2014wpa}.
Numerical studies can also include accreting mater to determine electromagnetic counterparts of the recoil \cite{Sijacki:2010tk,Ponce:2011kv,Meliani:2016rwn}.

\begin{figure}
\includegraphics[angle=0,width=0.95\columnwidth]{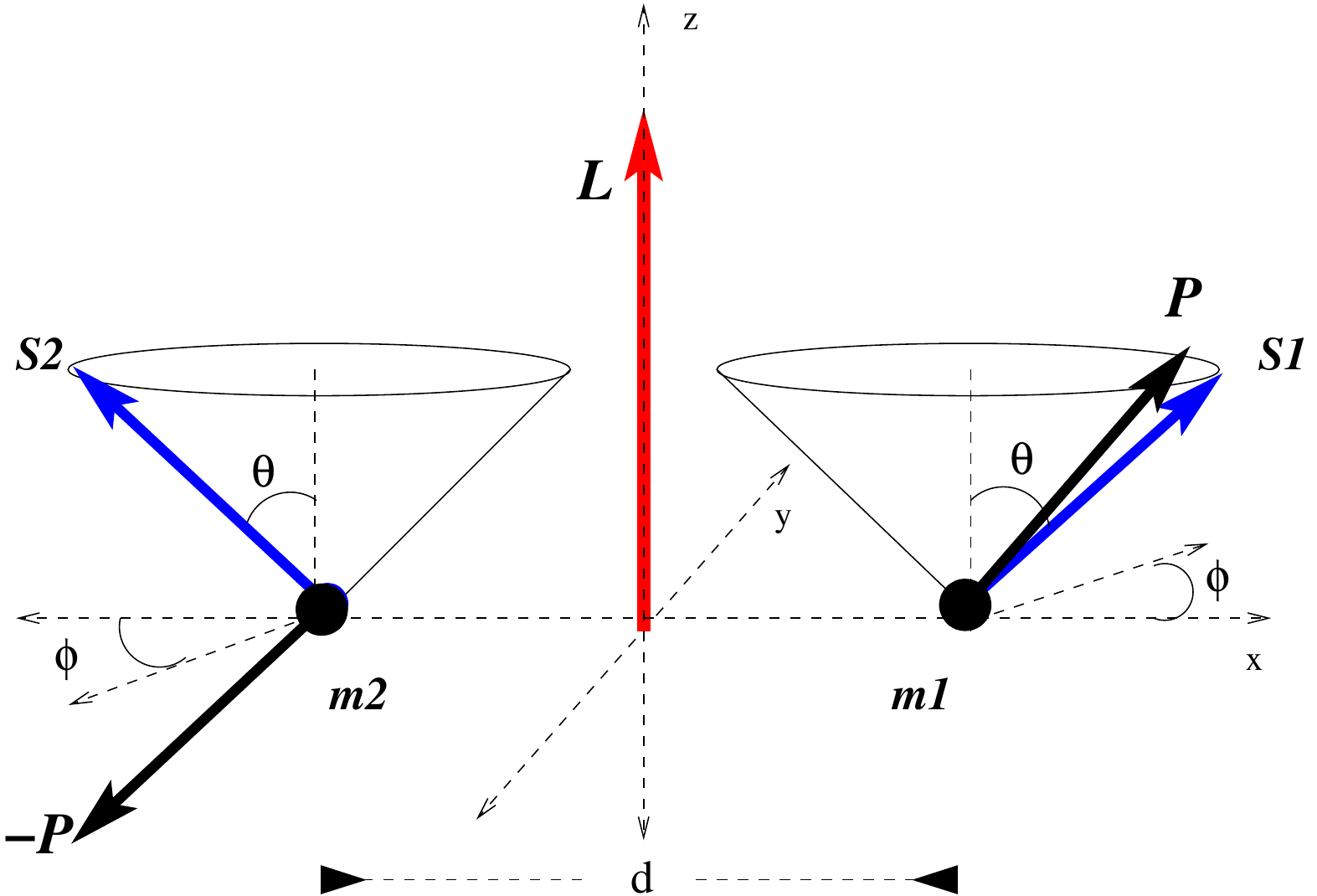}\\
\caption{The hangup-kick configurations to maximize the merger recoils. In our case spin magnitudes $S_{1,2}/m_{1,2}^2=0.97$, polar angle $\theta$ of $50.98^\circ$ and $m_1=m_2$. Simulations start at an initial coordinate separation $d/m=9$ and different angles $\varphi$ between the line connecting the BHs and the projection of the spin onto the orbital plane. 
\label{fig:hangupkick}}
\end{figure}

Interestingly, the observability of these recoils with gravitational wave detectors \cite{Gerosa:2016vip,CalderonBustillo:2018zuq} has been explored recently. Here we test this question in the most favorable scenario, that of the hangup-kick recoil with explicit simulations of nearly maximal spins ($\alpha=|\vec{\alpha}_{1,2}|=|\vec{S}_{1,2}|/m^2_{1,2}=0.97$). We compare waveforms for configurations within the hangup-kick family (See Fig. \ref{fig:hangupkick}) leading to nearly maximally but opposed recoils and passing through essentially vanishing recoil to see the required signal-to-noise ratio to distinguish between them with the analytic advanced LIGO sensitivity curve \cite{AdvLIGONC}. 

This paper is organized as follows, in the next section \ref{sec:nr} we describe
the numerical relativity techniques that we will use in the evolutions of the binary black holes. In section \ref{sec:sim} we describe the results of the simulations within the hangup family with equal mass black holes and spin magnitudes $\alpha=0.97$ for eight different spin orientations. This systematic study provides a method to fit a sinusoidal dependence of the recoil velocity of the final black hole as a function of the spin orientation. A new technique to identify this relative spin orientation at merger from the waveform phase is described in subsection
\ref{sec:peak}. We also analyze in subsection \ref{sec:convergence} the finite difference errors of our simulations by studying one member of the family with three resolutions and assess the differences with respect to its extrapolated value. We end the paper with a discussion, in section \ref{sec:dis}, of the potentially observable recoil effects on these waveforms. We evaluate the matching 
of our simulations with each other, taking as a reference the one with the lowest recoil velocity, to see the signal-to-noise (SNR) requirements to distinguish one from the other by advanced LIGO. We also come back to the first gravitational wave event GW150914, that we recently reanalyzed in Ref. \cite{Healy:2019jyf}, to evaluate the likelihood of recoils within a different simulation family, involving one single spinning black hole.

\section{Numerical Techniques}\label{sec:nr}

The late orbital dynamics of spinning binary black holes remain a
fascinating area of research since the 
numerical breakthroughs \cite{Pretorius:2005gq, Campanelli:2005dd, Baker:2005vv}
solved the binary black hole problem via supercomputer simulations.
Among the notable spin effects (without Newtonian analogs)
observed in supercomputer simulations are the 
hangup effect~\cite{Campanelli:2006uy},
which prompts or delays the merger of binary black holes depending on
the spin-orbit coupling, $\vec{S}\cdot\vec{L}$, being positive or negative
(aligned spins or antialigned spins with the orbital angular momentum $\vec{L}$);
the flip-flop of individual black hole spins passing from aligned to antialigned
periods with respect to the orbital angular momentum \cite{Lousto:2015uwa}
the alignment instability \cite{Kesden:2014sla}
as a case of imaginary flip-flop frequencies \cite{Lousto:2016nlp};
and the total flip of the orbital angular momentum \cite{Lousto:2018dgd}
under generic retrograde orbits for intermediate mass ratio binaries $(q<1/4)$.

Perhaps one of the most notable predictions of numerical relativity are the large
recoils (thousands of km/s) of the final black hole remnant
\cite{Campanelli:2007ew}, and up to 5,000 km/s \cite{Lousto:2011kp}.
Those results have been obtained from simulations with spinning black
holes of $\alpha=S/m^2=0.7, 0.8, 0.9$ and extrapolated to maximally spinning
holes. More recently, we introduced a version of
highly-spinning initial data, based on the superposition of two
Kerr black holes~\cite{Ruchlin:2014zva, Healy:2015mla},
in a puncture gauge that can easily be incorporated into moving-punctures
codes. In Refs.~\cite{Ruchlin:2014zva,Zlochower:2017bbg}, 
we were able to evolve an
equal-mass binary with aligned spins, and spin magnitudes of
$\alpha=0.95$ and $\alpha=0.99$ respectively, using this new data and compare 
with the SXS results of Ref.~\cite{Mroue:2013xna}, finding excellent
agreement.

In order to verify the extrapolation to near maximally spinning 
black holes and its evolution for a precessing  system
(in particular here the binary has a bobbing motion), we designed
a set of 8 simulations in the hangup-kick configuration with spins
$\alpha=0.97$. These simulations in turn will allow us to single out
the effect of recoil as a function of its merger phase and their
observability with gravitational wave detectors.

In table \ref{tab:ID} we provide the 8 configurations spanning
different initial orientations of the spin projection onto the orbital plane
$S_\perp$, with respect to the
line joining each hole as described by the angle $\varphi$, and are
chosen to include near maximum recoil in both z-directions ($\vec{L}$)
and near zero recoil.
Here $\varphi=\phi(t=0)$ and at that initial time $S_x=S_\perp\cos\varphi$ and
$S_y=S_\perp\sin\varphi$ for one black hole and reversed signs for the other.
The polar angle $\theta$ of the spin with respect to the orbital angular
momentum $\vec{L}$ has been chosen to maximize the recoil according
to the  predictions in Ref. \cite{Lousto:2012su}, i.e. reproduced here in Eqs.
(\ref{eq:FS}), (\ref{eq:pade}); and evaluated for $\alpha=0.97$ give the value
$\theta_{max}=50.98$ degrees.

We have chosen the initial separations of the binaries to guarantee
around 7 orbits before merger and in order to
estimate the accuracy of the finite resolution used in those simulations
we perform three simulations for a member of the family (that with
$\varphi=291^\circ$), at increasing resolutions by a factor 1.2 in order to study
the convergence of the relevant quantities for our studies. Those
results are reported later in subsection \ref{sec:convergence}.

\begin{table}
  \caption{Initial simple proper distance and spins of the BHs.  The initial 
coordinate separation in all cases is $D/m=9$ and mass ratio $q=1$.  ADM masses are between 0.9880 and 0.9884. 
Each simulation in the
series can be uniquely described by the azimuthal spin angle, $\varphi$.  Both spins have
magnitude 0.97 and polar angle $\theta$ of $50.98^\circ$, and the angle
$\Delta\varphi$ between $\vec \alpha_1$ and $\vec \alpha_2$
is $180^\circ$.  In terms of the spin components 
$\vec \alpha_1 = ( \alpha_{1x}, \alpha_{1y}, \alpha_{1z} ) = ( -\alpha_{2x}, -\alpha_{2y}, \alpha_{2z} )$.
\label{tab:ID}}
\begin{ruledtabular}
  \begin{tabular}{lccccc}
$\varphi$ & $m\Omega_{22}$ & $d/m$ & $\alpha_{2x}$ & $\alpha_{2y}$ & $\alpha_{2z}$ \\
\hline
0   & 0.01032 & 12.5183 &  0.7536 &  0.0000 & 0.6107 \\
30  & 0.01044 & 12.4045 &  0.6527 &  0.3768 & 0.6107 \\
60  & 0.01053 & 12.2011 &  0.3768 &  0.6527 & 0.6107 \\
90  & 0.01055 & 12.2128 &  0.0000 &  0.7536 & 0.6107 \\
120 & 0.01051 & 12.3744 & -0.3768 &  0.6527 & 0.6107 \\
150 & 0.01046 & 12.4913 & -0.6527 &  0.3768 & 0.6107 \\
203 & 0.01046 & 12.4455 & -0.6953 & -0.2908 & 0.6107 \\
291 & 0.01029 & 12.3250 &  0.2663 & -0.7050 & 0.6107 \\
\end{tabular}
\end{ruledtabular}
\end{table}



We evolve the binary black hole data sets using the {\sc
LazEv}~\cite{Zlochower:2005bj} implementation of the moving puncture
approach~\cite{Campanelli:2005dd} with the conformal
function $W=\sqrt{\chi}=\exp(-2\phi)$ suggested by
Ref.~\cite{Marronetti:2007wz}.  For the runs presented here, we use
centered, eighth-order finite differencing in
space~\cite{Lousto:2007rj}, a fourth-order Runge Kutta time
integrator, and a 7th-order Kreiss-Oliger dissipation operator.
We use a Courant factor of 0.25 in the CCZ4 formulation of
the evolution equations \cite{Alic:2011gg}.
Our code uses the {\sc EinsteinToolkit}~\cite{Loffler:2011ay,
einsteintoolkit} / {\sc Cactus}~\cite{cactus_web} /
{\sc Carpet}~\cite{Schnetter-etal-03b} infrastructure.  The {\sc
Carpet} mesh refinement driver provides a
``moving boxes'' style of mesh refinement. In this approach, refined
grids of fixed size are arranged about the coordinate centers of both
holes.  The evolution code then moves these fine grids about the
computational domain by following the trajectories of the two
black holes.
We use {\sc AHFinderDirect}~\cite{Thornburg2003:AH-finding} to locate
apparent horizons.  We measure at it the mass and the
magnitude of the horizon spin using
the {\it isolated horizon} (IH) algorithm detailed in
Ref.~\cite{Dreyer02a} and as implemented in Ref.~\cite{Campanelli:2006fy}.
We measure radiated energy,
linear momentum, and angular momentum, in terms of the radiative Weyl
Scalar $\Psi_4$, using the formulas provided in
Refs.~\cite{Campanelli:1998jv,Lousto:2007mh}.
We extract the radiated 
energy-momentum at finite radius and extrapolate to $r=\infty$
with the
perturbative extrapolation described in Ref.~\cite{Nakano:2015pta}.  
For the radiated quantities, we use all modes up to and including
$\ell_{max}=6$.
Quasicircular (low eccentricity) initial orbital parameters
are computed using the 3rd. order post-Newtonian techniques described in~\cite{Healy:2017zqj}.

\section{Results}\label{sec:sim}

We summarize the results of our BBH evolutions in Table \ref{tab:remnant}
where the final black hole remnant properties and peak waveform luminosity
values are reported. The modeling of remnant mass and spin for precessing
binaries is given in Ref. \cite{Lousto:2013wta,Zlochower:2015wga} with
both quantities bearing a $\cos2\phi$-dependence for the current family of
simulations.
Here, we will particularly focus on the analysis of the recoil velocities
with regards to the predictions for those simulations with high spin
$(\alpha=0.97)$ from the extrapolation of previous fitting formulae
cfr. in equations (\ref{eq:FS}) or (\ref{eq:pade}).

\begin{table*}
  \caption{Final properties of the remnant BH.  The final mass $M_f/m$, final spin $\alpha_f$,
recoil velocity in km/s, and peak luminosity in ergs/s are given.  The number of orbits before
merger and time of peak luminosity are also given. $\Delta\phi$ representing the relative phases with respect to the $\varphi=0$ case.
\label{tab:remnant}}
\begin{ruledtabular}
  \begin{tabular}{lcccccccc}
$\varphi$ & $\Delta\phi_{peak}$ & $\Delta\phi_{traj}$ & $2N_{orbits}$ & $M_f/m$ & $\alpha_f$ & $V_{recoil}$ & peak Lum. & $t_{Hpeak}$ \\
\hline
0   &       0 &      0 & 14.0095 & 0.9251 & 0.8525 & -4014 & 5.603\e{56} & 860.5 \\
30  &   33.05 &  29.85 & 13.9915 & 0.9217 & 0.8461 & -4622 & 6.076\e{56} & 860.1 \\
60  &   65.99 &  77.16 & 13.9859 & 0.9200 & 0.8446 & -3882 & 6.228\e{56} & 859.2 \\
90  &   86.17 & 120.87 & 13.9689 & 0.9215 & 0.8496 & -1846 & 5.811\e{56} & 859.8 \\
120 &  106.44 & 143.48 & 13.9685 & 0.9244 & 0.8550 &   531 & 5.390\e{56} & 851.4 \\
150 &  142.53 & 160.00 & 14.0011 & 0.9260 & 0.8565 &  2553 & 5.326\e{56} & 857.3 \\
203 &  203.51 & 201.85 & 13.9950 & 0.9225 & 0.8475 &  4579 & 5.965\e{56} & 860.4 \\
291 &  264.09 & 320.23 & 13.9673 & 0.9245 & 0.8536 &   186 & 5.487\e{56} & 861.3 \\
\hline
\end{tabular}
\end{ruledtabular}
\end{table*}

In order to analyze the results of the present simulations, We can fit the recoil
to the form
\begin{equation}
V_{\rm rec} = V_{1}
\cos(\Delta\phi + \phi_1) + V_{3} \cos(3 \Delta\phi + 3 \phi_3),
  \label{eq:phifit}
\end{equation}
where $V_1$, $V_3$, $\phi_1$, and $\phi_3$ are fitting parameters and $\Delta\phi$ is the initial phase of the spin with respect to a reference direction (in our case the y-axis).


Based on \cite{Lousto:2011kp}, we expected that
the recoil would have the form
\begin{eqnarray}
  V_1 =&& \left(V_{1,1} + V_A \alpha  \cos \theta
+ V_B \alpha^2 \cos^2\theta  + V_C \alpha^3 \cos^3\theta
\right)\times\nonumber \\ 
  &&\ \alpha \sin \theta,
  \label{eq:FS}
\end{eqnarray}
where $V_1$ is the component of the recoil proportional to $\cos
\phi$,  $V_{1,1}$ arises from the ``superkick'' formula, and the remaining
terms are proportional to linear, quadratic, and higher orders in
$S_z/m^2=\alpha \cos \theta$ (the spin component in the direction of
the orbital angular momentum).

A fit of the simulations reported in \cite{Lousto:2012su} to this ansatz~(\ref{eq:FS})
showed that the truncated series appears to converge very slowly
with coefficients $V_{1,1}=(3677.76\pm15.17)\ \KMS$,
$V_A=(2481.21\pm67.09)\ \KMS$, $V_B=(1792.45\pm92.98)\ \KMS$,
$V_C=(1506.52\pm286.61)\ \KMS$ that have
relatively large uncertainties. In what follows we will neglect
the contribution of $V_3\sim100$km/s; see \cite{Lousto:2012su} for more details.

In addition, we proposed the alternative modeling
\begin{equation}
  \tilde{V}_1 = \left(\frac{1 + E \alpha \cos
\theta}{1+F \alpha \cos \theta}\right)\,\, D \alpha \sin \theta
\label{eq:pade}
\end{equation}
which can be thought of as a resummation of Eq.~(\ref{eq:FS}) with an
additional term $E \alpha \cos \theta$,
and fit to $D$, $E$, $F$ (where we used the prediction
of~\cite{Lousto:2010xk} to model the  $\tilde{V}_1$ for $\theta=90^\circ$)
and found $D=(3684.73\pm5.67)\ \KMS$, $E=0.0705\pm0.0127$, and
$F=-0.6238\pm0.0098$. Note that $E$ is approximately $1/9$ of $F$,
indicating that coefficients in this series get progressively smaller
in a faster sequence than in Eq.~(\ref{eq:FS}).

We can fit to the recoil dependence on the initial angle $\phi$ between the
spin and the y-axis. Alternatively, one can seek a reference frame, closer to merger, when most of the net recoil appears to be generated.
In Refs. \cite{Lousto:2012gt,Zlochower:2015wga} we have described in
 a totally coordinate based frame (punctures trajectories) the way to extract
an instantaneous orbital plane and spin projections as displayed in
Figure 3 of reference \cite{Lousto:2012gt} or Figure 1 of \cite{Zlochower:2015wga}.
We implement here a new measure of this angle about merger with respect to the $\varphi=0$ case as
a reference. We introduce the notion of using the peak amplitude phase of the waveform $\phi_{peak}$,
as a measure for the reference phase of the recoil modeling and provide more
detail in subsection \ref{sec:peak}.

\begin{figure}
    \includegraphics[angle=270,width=\columnwidth]{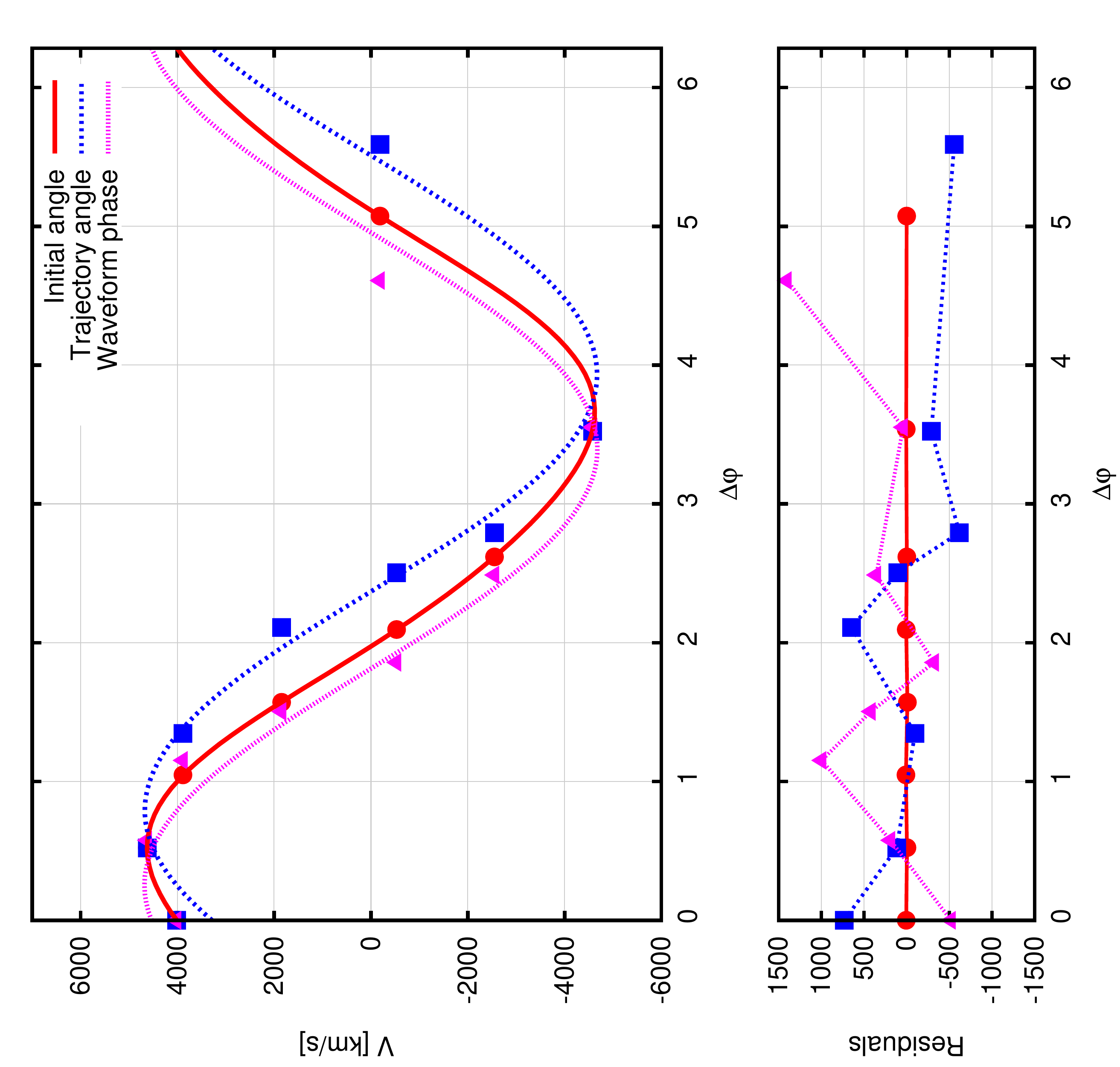}
    \caption{The plots show the fits for the three different evaluations of $\Delta\phi=\varphi=$ initial angle (in red circles),  $\Delta\phi_{traj}=$ trajectory angle as defined in \cite{Lousto:2012gt} (in magenta triangles), and  $\Delta\phi_{peak}$ from the waveform phase at the peak amplitude (in blue squares), as defined in this paper.
\label{fig:Vvsang}}
\end{figure}

These fits are displayed in Fig. \ref{fig:Vvsang} giving rise to an estimate of the maximum
recoil for these configurations in the form of the fitted amplitude of the sinusoidal dependence on $\Delta\phi$ as given by Eq. (\ref{eq:phifit}) to extract the leading $\cos\Delta\phi$-dependence and have a control of the nonleading $\cos3\Delta\phi$ term.
The three different evaluations of $\Delta\phi=\varphi=$ initial angle (in red circles),  $\Delta\phi_{traj}=$ trajectory angle as defined in \cite{Lousto:2012gt} (in magenta triangles), and  $\Delta\phi_{peak}$ from the waveform phase at the peak amplitude (in blue squares), as defined in subsection \ref{sec:peak} below.

Table \ref{tab:Vfit1} displays the measured fitting parameters and its statistical asymptotic standard
errors with 4 degrees of freedom.
Evaluating Eqs. (\ref{eq:FS}) and (\ref{eq:pade}) with the parameters for 
the series studied here ($\alpha=0.97$ and $\theta=50.98^\circ$), 
we find $V_1 = 4675.97 \pm 64.71$ and $\tilde{V}_1 = 4678.90 \pm 57.52$ km/s respectively.  
Comparing to the three fits given in table \ref{tab:Vfit1}, we see excellent 
agreement when using $\Delta\phi_{traj}$ ($4678.96 \pm 40.82$ km/s) 
and $\Delta\phi_{peak}$ ($4678.90 \pm 57.52$ km/s). 

\begin{table*}
  \caption{A fit $A_1\cos(\Delta\phi-\phi_1) + A_3\cos(3[\Delta\phi-\phi_3])$ to the $V_{recoil}$ with 4 degrees of freedom.
    For the three different evaluations of $\Delta\phi=\varphi=$ initial angle (in red circles),  $\Delta\phi_{traj}=$ trajectory angle as defined in \cite{Lousto:2012gt} (in magenta triangles)), and  $\Delta\phi_{peak}$ from the waveform phase at the peak amplitude (in blue squares), as defined in this paper.
\label{tab:Vfit1}}
\begin{ruledtabular}
  \begin{tabular}{l|cll|cll|cll}
Parameters & Initial angle & Standard & Error & Trajectory angle & Standard & Error & Waveform phase &  Standard & Error \\
\hline
$A_1$    & 4569.47 & $\pm 3.825$    & (0.083\%) & 4678.96  & $\pm  408.2$   & (8.724\%) & 4678.88 & $\pm  513.0$   & (10.96\%)\\
$\phi_1$ & 0.4353  & $\pm 0.0008 $  & (0.074\%) & 0.7960   & $\pm  0.0731 $ & (9.432\%) & 0.2447  & $\pm  0.1094 $ & (8.253\%)\\
$A_3$    & 152.22  & $\pm 3.822$    & (2.511\%) & 10.0268  & $\pm  388.4  $ & (3873\%)  & 9.96288 & $\pm  551.1  $ & (5531\%)\\
$\phi_3$ & 0.8814  & $\pm 0.00840$  & (0.147\%) & 0.0617   & $\pm  12.1   $ & (741\%)   & 0.7434  & $\pm  16.55  $ & (883\%)\\
\end{tabular}
\end{ruledtabular}
\end{table*}

\subsection{Reference frame at peak waveform amplitude}\label{sec:peak}

\begin{figure}
  \includegraphics[angle=270,width=\columnwidth]{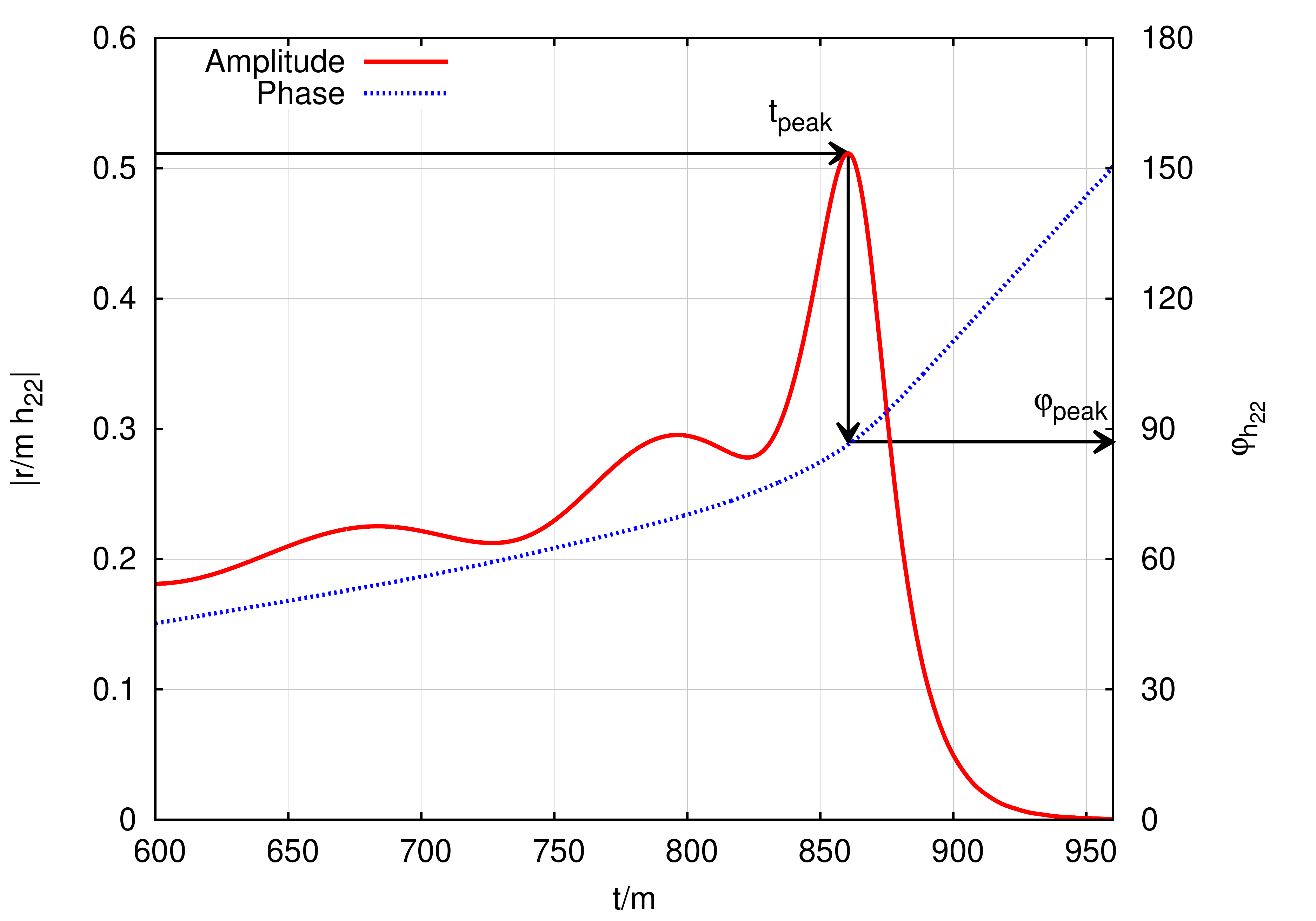}\\
    \caption{This figure displays the use of the method to determine the waveform phase at the peak amplitude from the simulation time at which this peak is observed for the $\varphi=0^\circ$ configuration. 
\label{fig:PeakL}}
\end{figure}

The peak amplitude $h_{peak}^{22}$ and peak waveform
frequency $\Omega_{peak}^{22}$
modeling in aligned binaries simulations was introduced in
Ref.~\cite{Healy:2018swt}. Here we use its definition to determine a
reference time and hence phase of the waveform at which we can assign a
recoil dependence of the form (\ref{eq:phifit}) and as represented in Fig. \ref{fig:PeakL}.
We compare this gauge invariant method to determine the differential (near merger) phase dependence to the
coordinate based method of \cite{Lousto:2012gt,Zlochower:2015wga} that was used in the original hangup-kick work \cite{Lousto:2011kp} and to determine the
numerical coefficients in Eqs. (\ref{eq:FS}) and (\ref{eq:pade}). Note that the two methods defined using a (near merger) measure as reference lead to very similar results. The statistical errors of those methods appear much larger than those measured from the initial angle $\phi$ given the difficulties in measuring directions in the strong dynamical regime of the merger phase.

The notion that the phase of the waveform at peak luminosity as a reference
in the strong field regime, near the merger of the two black holes, is a very
interesting one, since it is amenable to be generalized in the fully precessing case. In that case one has to determine the orbital plane orientation from
the direction of the maximum power of gravitational waves at the peak luminosity. Also measure the phase of the waveform along that privileged direction.
Appropriate families of simulations to extract modeling parameters should
then be designed. This will be the subject of a future research by the authors.

\begin{figure}
  \includegraphics[angle=270,width=\columnwidth]{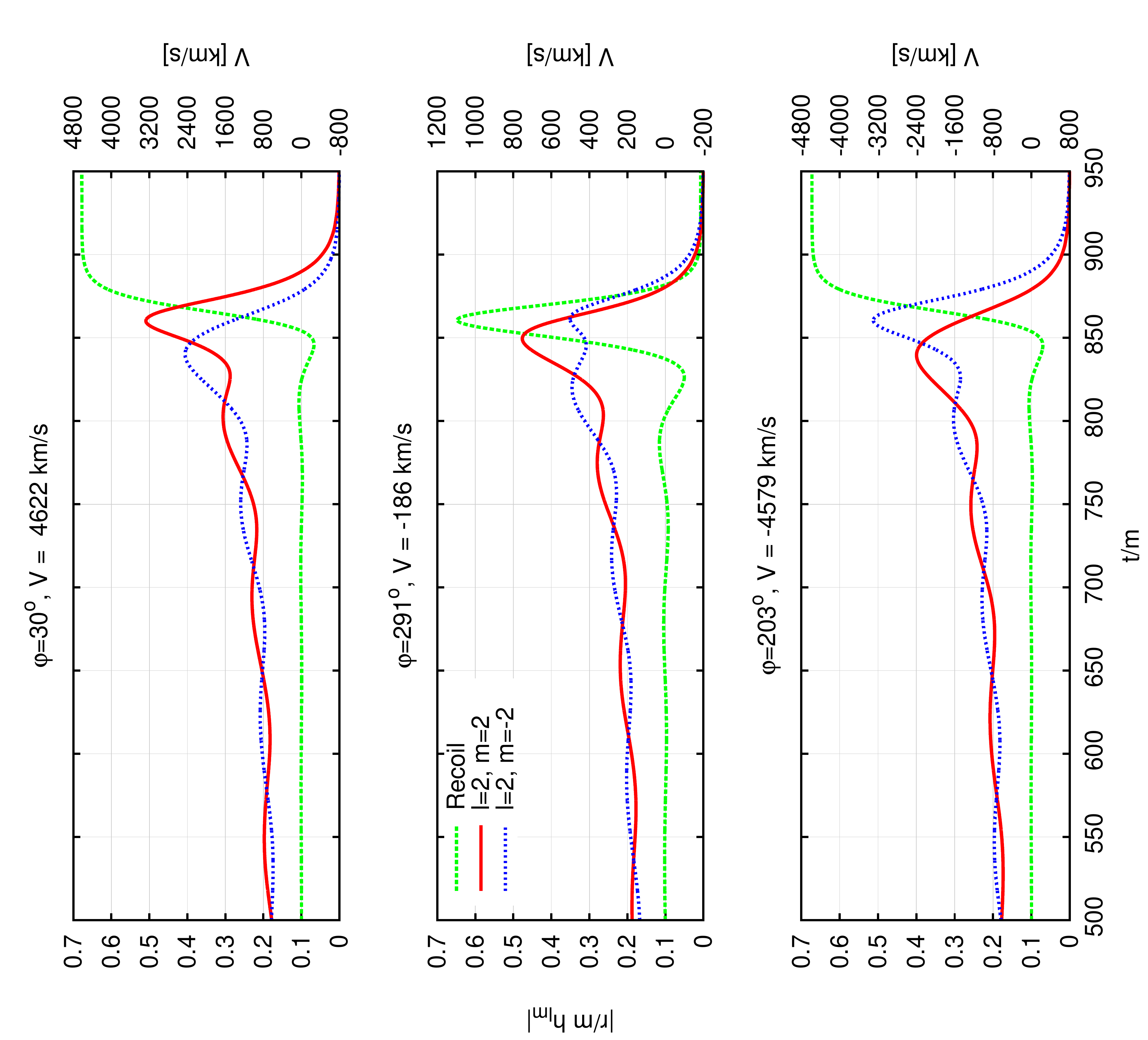}\\
    \caption{ Plots of the (2,2) (red) and (2,-2) (blue) modes of the strain for the three simulations
in the series that show near maximal, near minimal, and near zero recoil.  The recoil velocity versus time (green) is shown using the right $y$-axis. Note that in the bottom panel, the range of the right-hand $y$-axis is reversed (runs from +800 at the bottom to -4800 at the top).
\label{fig:A22A2m2}}
\end{figure}


\subsection{Recoil Generation}\label{sec:generation}
These systems provide an illustrative example of how the recoil is cumulated during
late inspiral, merger, and ringdown.  Due to the symmetry of these systems, the recoil
of the remnant BH is solely in the $z$-direction, which is aligned with
the gravitational wave extraction frame.  The recoil can be calculated from individual modes
of $\Psi_4 = \sum_{l=2}^\infty \sum_{m=-l}^l A^{l,m} \left( {}_{-2}Y^{l,m} (\theta,\phi) \right)$ by Eqs. (3.15), (3.18), and (3.19) in \cite{Ruiz:2007yx}:

\begin{eqnarray}
\frac{d P_z}{dt} &=& \lim_{r \to \infty} \frac{r^2}{16 \pi} \sum_{l,m}
\int_{-\infty}^t  dt' A^{l,m}\\
&\times& \int_{-\infty}^t dt' \: \Big( c_{l,m}\, \bar{A}^{l,m}
+ d_{l,m}\, \bar{A}^{l-1,m} \nonumber\\
&+&  d_{l+1,m}\, \bar{A}^{l+1,m} \Big) \; , \nonumber \label{eq:kz} \\
c_{l,m} &=& \frac{2\,m}{l\,(l+1)} \; , \nonumber\\
d_{l,m} &=& \frac{1}{l}\, \sqrt{\frac{(l-2)\,(l+2)\,(l-m)\,(l+m)}
{(2l-1)(2l+1)}} \; .\nonumber
\end{eqnarray}

Table \ref{tab:modes} shows the contributions to the recoil from the mode pairs of Eq. \ref{eq:kz}
that contribute more than 10 km/s for the three simulations that appear in Fig. \ref{fig:A22A2m2}.  These
three simulations are the ones with the near maximal, near zero, and near minimal recoil
velocities (top to bottom).  To good approximation, when the amplitude of the (2,2) mode 
is larger than the amplitude of the (2,-2) mode, the recoil velocity will increase.  This 
is easiest to see near merger, as in the top panel of Fig. \ref{fig:A22A2m2}, but is true
throughout.  In this panel, the red (2,2) dominates from late inspiral through ringdown,
resulting in a near maximal recoil for these configurations.  In the bottom panel, the
opposite is true, the blue (2,-2) dominates over the same range, and the recoil is approximately
the same, but in the opposite direction (note the $y$-axis on the right is reversed).  
The middle panel is interesting, in that it exhibits
a late-time continuation of the orbital wobbling leading to an in-phase cancellation or 
anti-kick, where at first we obtain a large recoil (around 1,000 km/s) followed by another
large recoil which cancels the original, resulting in a final recoil close to 0.  
This anti-kick can be explained again by which mode is dominating near merger.  At first, the
blue (2,-2) is dominating in the late inspiral, but as we approach the peak, red (2,2) dominates,
producing the large positive recoil.  However, during ringdown, blue (2,-2) dominates again
producing the large negative recoil cancellation.  Table \ref{tab:modes} shows that the
contributions of the (2,2) and (2,-2) mode with themselves produce the largest contributions
to the recoil, but will always carry an opposite sign (because of the $c_{lm}$ coefficient.)
For the near maximal and near minimal configurations, these two modes account for approximately
$90\%$ of the kick, leaving the remaining approximately 400 km/s to the other mode pairs.
Interestingly in the near zero configuration, the (2,2) and (2,-2) mode pairs only contribute 
85 km/s after the cancellation, leaving the bulk of the recoil (an additional 100 km/s) to the
higher mode pairs.  If the same analysis were applied to an aligned system, where the spins
are aligned with the orbital angular momentum, we would still obtain very large recoil
contributions from the (2,2) and (2,-2) mode pairs.  However, due to the symmetry, these would cancel
completely (and all other mode pairs), to give a net-zero recoil in the $z$-direction.

\begin{table}
  \caption{Mode pair contributions to the recoil velocity in the $z$-direction as in 
Eq. \ref{eq:kz} for the near maximal, near zero, and near minimal recoil configurations.
Only pairs with contributions $>10$ km/s are included here.
\label{tab:modes}}
\begin{ruledtabular}
  \begin{tabular}{rrrrrrr}
$\ell_1$ & $m_1$ & $\ell_2$ & $m_2$ & $V(\varphi=30^\circ)$ & $V(\varphi=291^\circ)$ & $V(\varphi=203^\circ)$ \\
\hline
2 &  2 & 2 &  2 &  9122.37 &  6779.79 &  4818.65 \\
2 & -2 & 2 & -2 & -4893.59 & -6865.65 & -9019.23 \\
2 & -2 & 3 & -2 &  -228.74 &  -435.80 &  -507.18 \\
2 &  2 & 3 &  2 &   521.46 &   334.62 &   227.70 \\
3 &  2 & 3 &  2 &    26.51 &    14.06 &    10.35 \\
3 & -2 & 3 & -2 &   -10.09 &   -25.43 &   -25.50 \\
4 &  4 & 4 &  4 &    85.99 &    47.51 &    20.80 \\
4 & -4 & 4 & -4 &   -21.93 &   -32.93 &   -84.98 \\
\hline
\end{tabular}
\end{ruledtabular}
\end{table}

\subsection{Convergence of the numerical simulations}\label{sec:convergence}

Numerous convergence studies of our past simulations have been performed.
In Appendix A of Ref.~\cite{Healy:2014yta},
in Appendix B of Ref.~\cite{Healy:2016lce}, and
for nonspinning binaries are reported in Ref.~\cite{Healy:2017mvh}.
For very highly spinning black holes ($s/m^2=0.99$)
convergence of evolutions was studied in Ref. \cite{Zlochower:2017bbg}
and for ($s/m^2=0.95$) in Ref. \cite{Healy:2017vuz} for unequal mass binaries.
For our current simulations we studied in detail
one member of the hangup kick family,
that with the lowest recoil, at an initial spin orientation angle $\phi=291^\circ$.
With three resolutions, lowering our standard resolution for the whole family
by factors of 1.2. Resolutions are denoted by ``NXXX", 
where XXX is a number related to the resolution in the wavezone.  For example,
``N144", the standard resolution for these simulations, has a wavezone resolution of M/1.44, and ``N100", has a resolution of M/1.00.  We then assume that
a quantity $\Psi$ behaves with resolution $h$ in the range of low $h_L$ to high $h_H$ as $\Psi(h)=\Psi(\text{exact}) + A\, h^n$, where we compute $\Psi(h)$ at the three resolutions $h_L, h_M, h_H$.
We evaluate the extrapolation to infinite resolution $\Psi(\text{exact})\to\Psi_\infty$ as
\bes
\beq
\Psi_{{\infty }}={\frac {\Psi_{{H}}\Psi_{{
        L}}-{\Psi_{{M}}}^{2}}{\Psi_{{H}}-2\,\Psi_{{M}}+\Psi_{{L}}}}
\eeq
\beq
A={\frac {{\Psi_{{L}}}^{2}-2\,\Psi_{{L}}\Psi_{{M}}+{\Psi_{{M}
    }}^{2}}{\Psi_{{H}}-2\,\Psi_{{M}}+\Psi_{{L}}}},
\eeq
\beq
n=-{\frac {1}{\ln 
 \left( f \right) }\ln  \left| {\frac {\Psi_{{H}}-\Psi_{{M}}}{\Psi_{
{L}}-\Psi_{{M}}}} \right| },
\eeq
\ees
where we also determine the constant $A$ and the convergence rate $n$.
We have also assumed that the low, medium, and high resolutions are related
by a common factor $f$ as $h_M=h_L/f$ and $h_H=h_L/f^2$,
as is the case presented here with $f=1.2$.

We found roughly the expected 4th-6th order convergence as displayed in Table \ref{tab:convergence} for the 
values of the recoil velocity and peak luminosity as well as
the final black hole mass and spin (as measured on
the apparent horizon.)
The results of an extrapolation to infinite resolution and the differences
with respect to the standard resolution (labeled as N144) are displayed in Table \ref{tab:convergence} to provide a measure of the expected errors for the
whole family of simulations.
Generically, for other simulations, we monitor the accuracy by measuring
the conservation of the individual horizon masses and spins during evolution
as well as the level of satisfaction of the Hamiltonian and momentum constraints.  
All eight N144 configurations show comparable behavior in these quantities.

\begin{table*}
  \caption{Convergence of key quantities for the $\varphi=291^\circ$ system with three resolutions. Richardson extrapolation
is used to determine the convergence order and infinitely extrapolated values.  Recoil velocities are given in km/s and peak
luminosities are ergs/s.  The fifth row shows the difference between the extrapolated and N144 values, and the sixth row 
shows the percent difference between the two.  There is an exception for the quantity in the last column, $\phi_{h_{22},\mathrm{peak}}$.  If we were to take the phase at a fixed time near peak for each resolution, we would observe and order of convergence between 5 and 6.  However, since we take the phase at the peak for each resolution, and the time of peak
is already convergent at an order of 5.5, we observe higher than normal convergence for the phase when measured this way.
\label{tab:convergence}}
\begin{ruledtabular}
  \begin{tabular}{lccccccc}
             & $V_{recoil}$ & $\alpha_f$ & $M_f/m$ & $10^{-56}\cdot\mathcal{L}_{\mathrm{peak}}$ & $|rh_{22}|_{\mathrm{peak}}$ & $t_{h_{22},\mathrm{peak}}$ & $\phi_{h_{22},\mathrm{peak}}$ \\
\hline
N100         & 227.41 & 0.853399 & 0.923310 & 5.4062 & 0.475254 & 962.804 & 89.793 \\
N120         & 193.35 & 0.853569 & 0.923599 & 5.4578 & 0.476050 & 962.595 & 89.800 \\
N144         & 186.03 & 0.853642 & 0.923705 & 5.4867 & 0.476328 & 962.519 & 89.801 \\
Inf. Extrap. & 184.03 & 0.853697 & 0.923766 & 5.5235 & 0.476476 & 962.476 & * \\
\hline
Inf. - N144  & -2.00 & 0.000055 & 0.000061  & 0.0368 & 0.000148 & -0.043  & *  \\
\% difference& -1.09 & 0.0065   & 0.0066    & 0.6673 & 0.0311   & -0.005  & *  \\
\hline
Conv. Order  &   8.4  & 4.6      & 5.5      & 3.2         & 5.8      & 5.5     & *   \\
\end{tabular}
\end{ruledtabular}
\end{table*}

\section{Discussion}\label{sec:dis}

We compute the waveforms $a$ and $b$ matching as the inner product in frequency $f$-domain
\begin{equation}
\label{eq:overlap}
\mathcal{M}=
\langle a|b\rangle_{k}\equiv 2 \int_{|f|>f_{\rm min}}df\frac{[\tilde{a}(f)]^{*}\tilde{b}(f)}{S_{h,k}(|f|)}.
\end{equation}
where the k$^{th}$ detector's noise power spectrum is $S_{h,k}(f)$
and we adopt a low-frequency cutoff $f_{\rm min}$.  By construction, we maximize over both a time and
phase shift between waveforms.
For our analysis of GW150914, we adopt the same noise power spectrum employed in previous work \cite{Abbott:2016apu,2018arXiv180510457L}, the advanced LIGO design sensitivity noise curve.  We use a reference total mass of $M_{total} = 74 M_{\odot}$ and $f_{\rm min}=30Hz$.  This choice of $M_{total}$ starts our waveform frequencies just below 30Hz after an initial windowing function is applied.
The minimal SNR needed to distinguish
between the two waveforms, given the mismatch
is $\mathrm{SNR}^2 \ge \frac{1}{1-\mathcal{M}}$.  

\begin{figure}
    \includegraphics[angle=270,width=0.95\columnwidth]{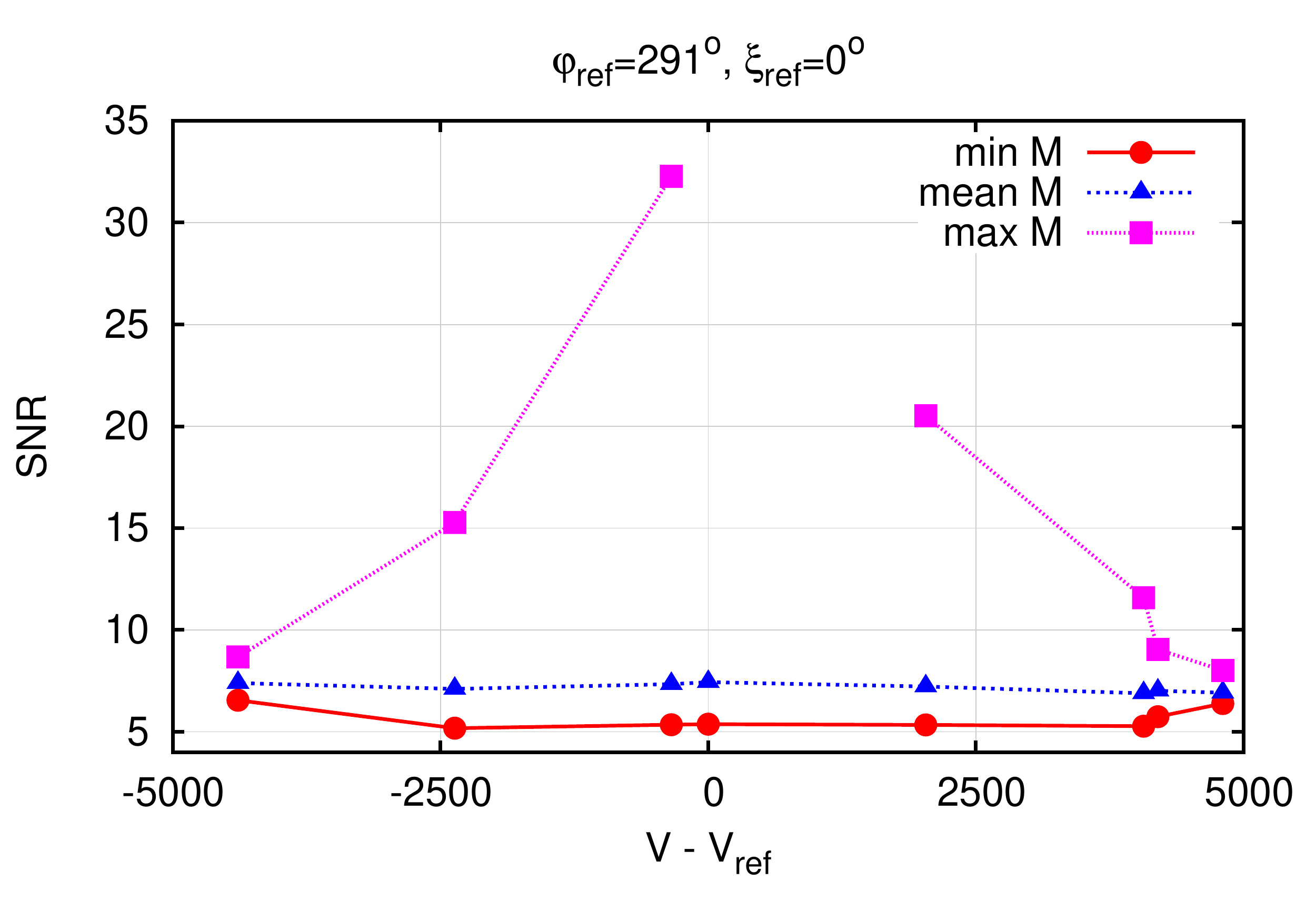}
    \caption{We display the SNR required to differentiate the hangup recoil waveforms from the member of the family with lowest recoil magnitude, $V_{ref}=186$ km/s at initial $\varphi=291^\circ$ Larger recoils are easier to mismatch, while nearby ones require SNR above 20. Lines indicate the minimum, maximum, and mean matches over the azimuthal and polar reconstruction angles of a given configuration against the $\varphi=291^\circ$, $\xi=0$ reference configuration.
\label{fig:snr}}
\end{figure}

To determine if waveforms from within this family of configurations can be distinguished between
different members of the family, we perform a series of matches between configurations.
That is, we choose a simulation and reconstruct the 
gravitational wave at a given polar and azimuthal angle and use this as our reference waveform.  For
each of the other configurations in the series, we can then calculate the match against our reference
waveform and produce a ``world map" of matches. We calculate the match
\begin{equation}
\label{eq:Mmap}
\mathcal{M}_i(\xi,\psi)=
\langle \varphi_{ref}[\xi_{ref},\psi_{ref}]|\varphi_i[\xi,\psi]\rangle,
\end{equation}
where $i$ runs over each configuration, and where $\xi$ and $\psi$ are the angles used
to reconstruct the second waveform at a given point in the skymap: 
$0\le\xi\le\pi$, and $-\pi<\psi\le\pi$.
In Fig. \ref{fig:snr}, we chose $\varphi_{ref}=291^\circ$ reconstructed at $\xi_{ref}=0^\circ=\psi_{ref}$ and calculate the SNR from the minimum, maximum, and mean matches over the world map.
We show that the last few cycles of the gravitational waveform from
black holes in the the hangup-kick configuration leading to a large
recoil of the final remnant of the BBH merger is potentially measurable by LIGO with reasonable SNR, i.e. around approximately 30. 
For comparison, the matching between different resolutions of the reference case, $\varphi=291^\circ$, gives us SNR of the order of 96 and 25 for N120 and N100 respectively. Extrapolation to infinite resolution of the simulations $\text{N}_\infty$ leads to a SNR of over 100 in order to differentiate the N144 from the $\text{N}_\infty$ result.

Given the spin misalignments of comparable masses BBH observed in the current detections \cite{LIGOScientific:2018mvr}, these kind of configurations seems not so unlikely to occur in nature.
While the search for detecting very highly spinning black holes with gravitational wave observations continues, it is important to search for them with the appropriated highly spinning templates and our simulations can contribute to fill in this gap near maximally spinning holes and properly cover this region of BBH parameter space. Parameter estimation techniques directly using numerical relativity waveforms from catalogs have been applied successfully for GW150914\cite{Healy:2019jyf} and GW170104\cite{Healy:2017abq} and will be the subject of further applications for O2 LIGO-Virgo observations. 

Phenomenological modeling of waveforms, such as the PhenomP \cite{Schmidt:2014iyl} mimic
precession from rotating aligned cases which leads to misevaluations of the recoil. See however new attempts to take recoil into account in other waveform models \cite{Chamberlain:2018snj,Gerosa:2018qay}. An improved analysis of GW150914 using a two spins effective one body model is provided in \cite{Abbott:2016izl}.

\begin{figure}
    \includegraphics[angle=270,width=0.95\columnwidth]{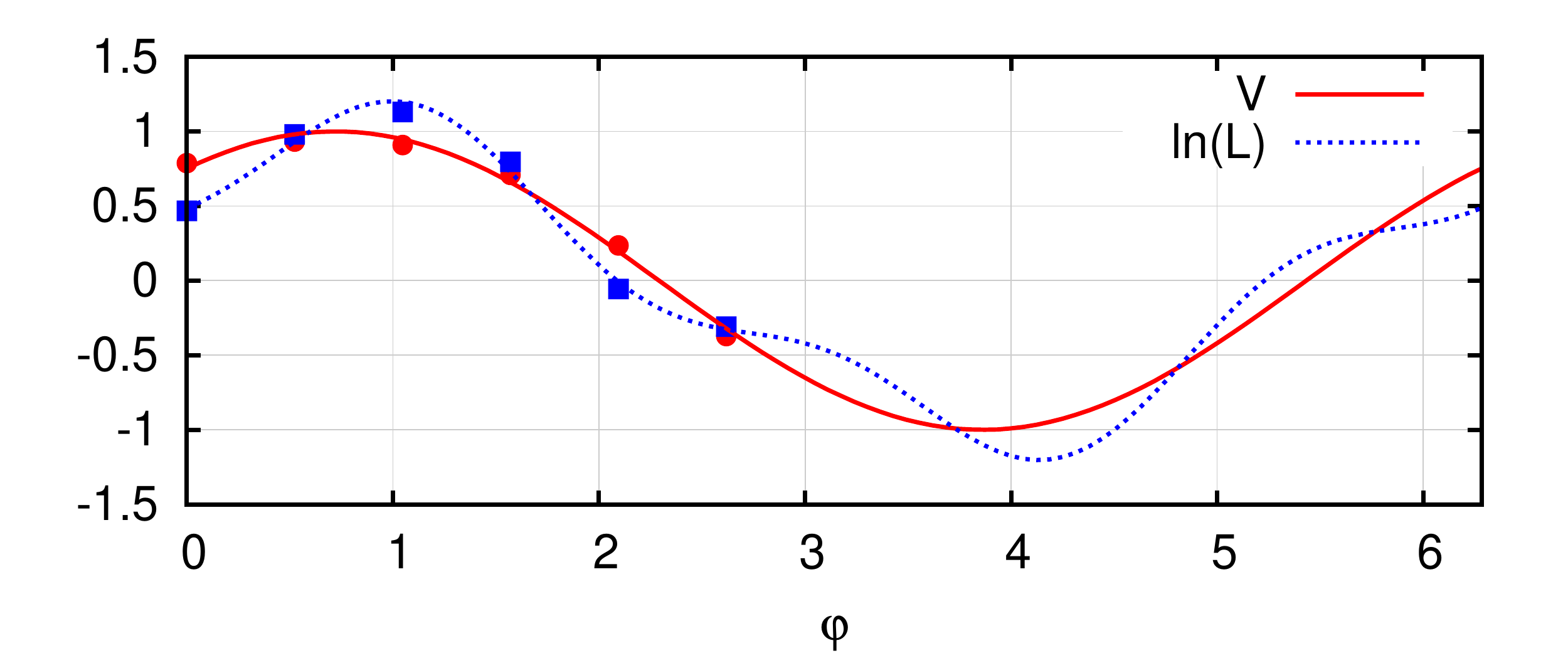}\\
    \includegraphics[angle=270,width=0.95\columnwidth]{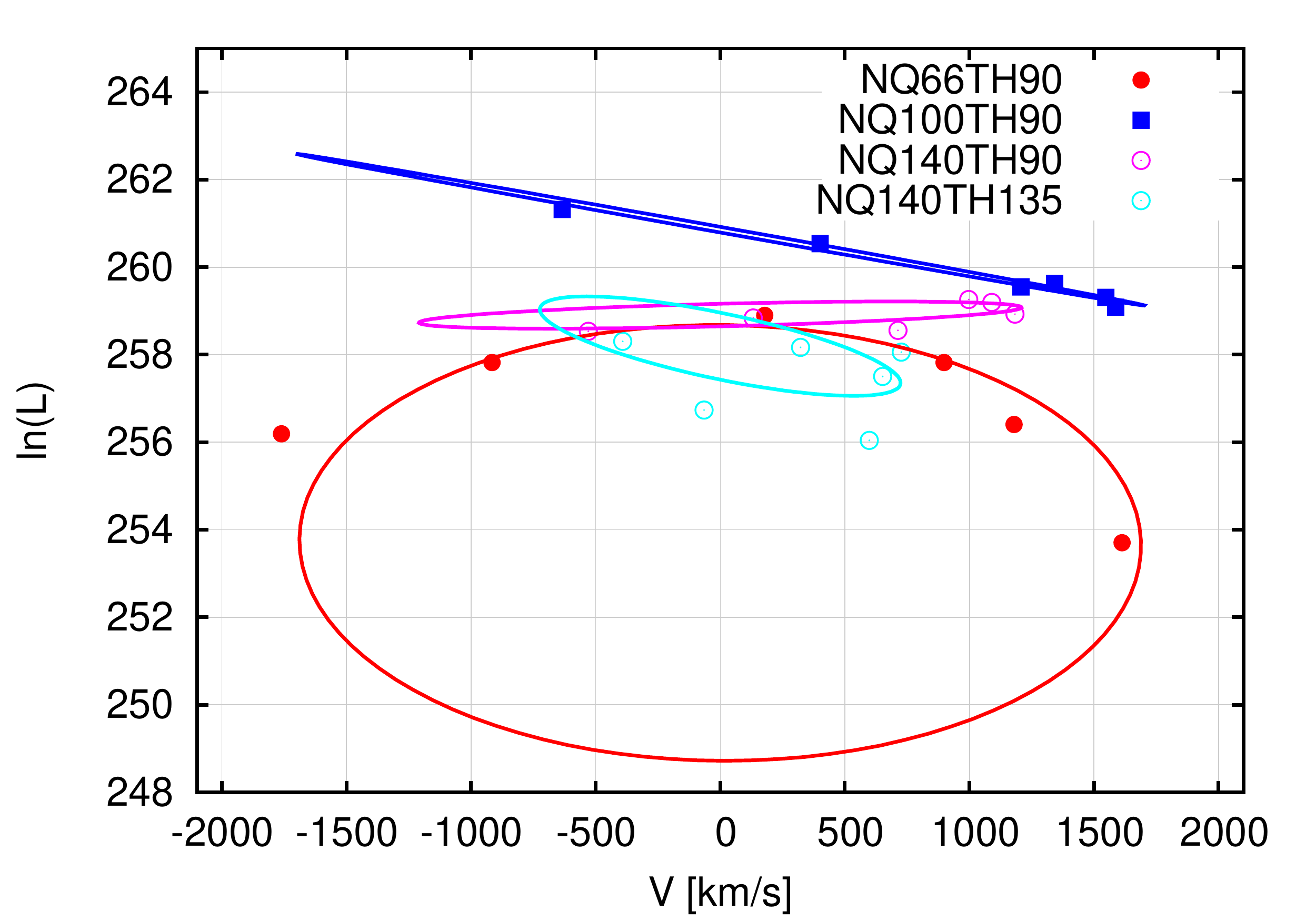}
    \caption{The orbits in the maximum likelihood $\ln {\cal L}$ and recoil velocity space of the simulations
      reported in Figure 8 of Ref. \cite{Healy:2019jyf} for GW150914.  Top panel shows the fit to the $\ln{\mathcal{L}}$ and recoil separately that together form the ellipses in the bottom panel. The recoil velocity uses a fit $V=A_1\cos(\varphi-\phi_1)$ and the log-likelihood uses $\ln{\mathcal{L}}=A_2\cos(\varphi-\phi_2) + A_3\cos(3\varphi-3\phi_3) + B$, where $A_1, \phi_1, A_2, \phi_2, A_3, \phi_3,$ and $ B$ are fitting parameters.  The data and ellipses in the bottom panel subtract off the $cos(3\varphi)$ terms. 
\label{fig:GW150914}}
\end{figure}

In Ref. \cite{Healy:2019jyf} we have been able to use a different family of simulations of binary black holes with one single spinning hole with amplitude $\alpha=0.8$ at all different orientations covering the two dimensional space of initial $(\theta,\varphi)$. Those lead to a ``world heat map'' as shown in the figure 8 of \cite{Healy:2019jyf} for the likelihood $\ln {\cal L}$ to represent the signal GW150914. Bit-equivalent data to the frames used for this study is available through GWOSC (Gravitational Wave Open Science Center) \cite{Vallisneri:2014vxa}, and the likelihood, $\ln {\cal L}$, is calculated using the RIFT framework \cite{2015PhRvD..92b3002P,Lange:2018pyp}. In addition to this 3-parameter space estimation, we can consider the subfamily with the mass ratio $q$ and inclination angle $\theta$ leading to the highest likelihood $\ln {\cal L}$ and use this one remaining $\phi$-parametrized subfamily to parametrize the $\phi$-dependence of the recoil. The resulting ``orbits'' from the interpolation of the data are displayed in Fig. \ref{fig:GW150914}, showing the top three $\ln {\cal L}$ families and the preference
for recoils of about $-1,500$ km/s.

Ultimately, determining accurately the recoil of the final hole from a binary system is paramount to determine (given a mass ratio) the spin orientations at merger.
Being able to determine the ``phase'' of the spin relative to the linear momentum of the holes at the merger (as determined by the maximum amplitude of radiation) allows to predict the recoil of the remnant black hole. Such determination has been performed for GW150914 \cite{Healy:2019jyf} 
leading to estimated recoils of around 1,500 km/s as displayed in Fig. \ref{fig:GW150914}. The differences this induces on the merger and ringdown phases can be estimated as well, as a consistency check and a test of the theory of gravitation.

For the source of GW150914
we were also able to estimate the inclination of the orbit from purely numerical waveforms, as displayed in
Figure 9  of Ref. \cite{Healy:2019jyf}. The ability to find a single maximum, not bimodal, orientation of the binary, is somewhat related to the measure of precession and this in turn is related to the spin misalignment with the orbital angular momentum that may induce large recoil velocities, those depending on the merger phase that we model in this paper for the maximum recoil configurations.

The application of this techniques that we tested in the case of the first gravitational wave signal GW150914, can be used in other detections of BBH mergers, as GW170104 and others in O2 \cite{LIGOScientific:2018mvr} and forthcoming observations and will be the subject of a future paper by the authors.

\begin{acknowledgments}
The authors thank R. O'Shaughnessy and Y. Zlochower for discussions on this work and H. Pfeiffer for comments on the original manuscript.
The authors gratefully acknowledge the National Science Foundation (NSF)
for financial support from Grants
 No.\ PHY-1912632, No.\ PHY-1607520, No.\ PHY-1707946, No.\ ACI-1550436, No.\ AST-1516150,
No.\ ACI-1516125, No.\ PHY-1726215.
This work used the Extreme Science and Engineering
Discovery Environment (XSEDE) [allocation TG-PHY060027N], which is
supported by NSF grant No. ACI-1548562.
Computational resources were also provided by the NewHorizons,
BlueSky Clusters, and Green Prairies
at the Rochester Institute of Technology, which were
supported by NSF grants No.\ PHY-0722703, No.\ DMS-0820923, No.\
AST-1028087, No.\ PHY-1229173, and No.\ PHY-1726215.
Computational resources were also provided by the Blue Waters sustained-petascale computing NSF projects OAC-1811228, OAC-0832606, OAC-1238993, OAC-1516247 and OAC-1515969, OAC-0725070. Blue Waters is a joint effort of the University of Illinois at Urbana-Champaign and its National Center for Supercomputing Applications.
This research has made use of data, software and/or web tools obtained from the Gravitational Wave Open Science Center (https://www.gw-openscience.org), a service of LIGO Laboratory, the LIGO Scientific Collaboration and the Virgo Collaboration. LIGO is funded by the U.S. National Science Foundation. Virgo is funded by the French Centre National de Recherche Scientifique (CNRS), the Italian Istituto Nazionale della Fisica Nucleare (INFN) and the Dutch Nikhef, with contributions by Polish and Hungarian institutes.
\end{acknowledgments}

\bibliographystyle{apsrev4-1}
\bibliography{../../Bibtex/references}

\begin{thebibliography}{86}%
\makeatletter
\providecommand \@ifxundefined [1]{%
 \@ifx{#1\undefined}
}%
\providecommand \@ifnum [1]{%
 \ifnum #1\expandafter \@firstoftwo
 \else \expandafter \@secondoftwo
 \fi
}%
\providecommand \@ifx [1]{%
 \ifx #1\expandafter \@firstoftwo
 \else \expandafter \@secondoftwo
 \fi
}%
\providecommand \natexlab [1]{#1}%
\providecommand \enquote  [1]{``#1''}%
\providecommand \bibnamefont  [1]{#1}%
\providecommand \bibfnamefont [1]{#1}%
\providecommand \citenamefont [1]{#1}%
\providecommand \href@noop [0]{\@secondoftwo}%
\providecommand \href [0]{\begingroup \@sanitize@url \@href}%
\providecommand \@href[1]{\@@startlink{#1}\@@href}%
\providecommand \@@href[1]{\endgroup#1\@@endlink}%
\providecommand \@sanitize@url [0]{\catcode `\\12\catcode `\$12\catcode
  `\&12\catcode `\#12\catcode `\^12\catcode `\_12\catcode `\%12\relax}%
\providecommand \@@startlink[1]{}%
\providecommand \@@endlink[0]{}%
\providecommand \url  [0]{\begingroup\@sanitize@url \@url }%
\providecommand \@url [1]{\endgroup\@href {#1}{\urlprefix }}%
\providecommand \urlprefix  [0]{URL }%
\providecommand \Eprint [0]{\href }%
\providecommand \doibase [0]{http://dx.doi.org/}%
\providecommand \selectlanguage [0]{\@gobble}%
\providecommand \bibinfo  [0]{\@secondoftwo}%
\providecommand \bibfield  [0]{\@secondoftwo}%
\providecommand \translation [1]{[#1]}%
\providecommand \BibitemOpen [0]{}%
\providecommand \bibitemStop [0]{}%
\providecommand \bibitemNoStop [0]{.\EOS\space}%
\providecommand \EOS [0]{\spacefactor3000\relax}%
\providecommand \BibitemShut  [1]{\csname bibitem#1\endcsname}%
\let\auto@bib@innerbib\@empty
\bibitem [{\citenamefont {Campanelli}\ \emph
  {et~al.}(2007{\natexlab{a}})\citenamefont {Campanelli}, \citenamefont
  {Lousto}, \citenamefont {Zlochower},\ and\ \citenamefont
  {Merritt}}]{Campanelli:2007ew}%
  \BibitemOpen
  \bibfield  {author} {\bibinfo {author} {\bibfnamefont {M.}~\bibnamefont
  {Campanelli}}, \bibinfo {author} {\bibfnamefont {C.~O.}\ \bibnamefont
  {Lousto}}, \bibinfo {author} {\bibfnamefont {Y.}~\bibnamefont {Zlochower}}, \
  and\ \bibinfo {author} {\bibfnamefont {D.}~\bibnamefont {Merritt}},\
  }\href@noop {} {\bibfield  {journal} {\bibinfo  {journal} {Astrophys. J.}\
  }\textbf {\bibinfo {volume} {659}},\ \bibinfo {pages} {L5} (\bibinfo {year}
  {2007}{\natexlab{a}})},\ \Eprint {http://arxiv.org/abs/gr-qc/0701164}
  {gr-qc/0701164} \BibitemShut {NoStop}%
\bibitem [{\citenamefont {Campanelli}\ \emph
  {et~al.}(2007{\natexlab{b}})\citenamefont {Campanelli}, \citenamefont
  {Lousto}, \citenamefont {Zlochower},\ and\ \citenamefont
  {Merritt}}]{Campanelli:2007cga}%
  \BibitemOpen
  \bibfield  {author} {\bibinfo {author} {\bibfnamefont {M.}~\bibnamefont
  {Campanelli}}, \bibinfo {author} {\bibfnamefont {C.~O.}\ \bibnamefont
  {Lousto}}, \bibinfo {author} {\bibfnamefont {Y.}~\bibnamefont {Zlochower}}, \
  and\ \bibinfo {author} {\bibfnamefont {D.}~\bibnamefont {Merritt}},\
  }\href@noop {} {\bibfield  {journal} {\bibinfo  {journal} {Phys. Rev. Lett.}\
  }\textbf {\bibinfo {volume} {98}},\ \bibinfo {pages} {231102} (\bibinfo
  {year} {2007}{\natexlab{b}})},\ \Eprint {http://arxiv.org/abs/gr-qc/0702133}
  {gr-qc/0702133} \BibitemShut {NoStop}%
\bibitem [{\citenamefont {Volonteri}(2007)}]{Volonteri:2007et}%
  \BibitemOpen
  \bibfield  {author} {\bibinfo {author} {\bibfnamefont {M.}~\bibnamefont
  {Volonteri}},\ }\href {\doibase 10.1086/519525} {\bibfield  {journal}
  {\bibinfo  {journal} {Astrophys. J.}\ }\textbf {\bibinfo {volume} {663}},\
  \bibinfo {pages} {L5} (\bibinfo {year} {2007})},\ \Eprint
  {http://arxiv.org/abs/astro-ph/0703180} {arXiv:astro-ph/0703180 [astro-ph]}
  \BibitemShut {NoStop}%
\bibitem [{\citenamefont {Loeb}(2007)}]{Loeb:2007wz}%
  \BibitemOpen
  \bibfield  {author} {\bibinfo {author} {\bibfnamefont {A.}~\bibnamefont
  {Loeb}},\ }\href {\doibase 10.1103/PhysRevLett.99.041103} {\bibfield
  {journal} {\bibinfo  {journal} {Phys. Rev. Lett.}\ }\textbf {\bibinfo
  {volume} {99}},\ \bibinfo {pages} {041103} (\bibinfo {year} {2007})},\
  \Eprint {http://arxiv.org/abs/astro-ph/0703722} {arXiv:astro-ph/0703722}
  \BibitemShut {NoStop}%
\bibitem [{\citenamefont {{Holley-Bockelmann}}\ \emph
  {et~al.}(2008)\citenamefont {{Holley-Bockelmann}}, \citenamefont
  {{G{\"u}ltekin}}, \citenamefont {{Shoemaker}},\ and\ \citenamefont
  {{Yunes}}}]{HolleyBockelmann:2007eh}%
  \BibitemOpen
  \bibfield  {author} {\bibinfo {author} {\bibfnamefont {K.}~\bibnamefont
  {{Holley-Bockelmann}}}, \bibinfo {author} {\bibfnamefont {K.}~\bibnamefont
  {{G{\"u}ltekin}}}, \bibinfo {author} {\bibfnamefont {D.}~\bibnamefont
  {{Shoemaker}}}, \ and\ \bibinfo {author} {\bibfnamefont {N.}~\bibnamefont
  {{Yunes}}},\ }\href {\doibase 10.1086/591218} {\bibfield  {journal} {\bibinfo
   {journal} {Astrophys. J.}\ }\textbf {\bibinfo {volume} {686}},\ \bibinfo
  {pages} {829} (\bibinfo {year} {2008})},\ \Eprint
  {http://arxiv.org/abs/0707.1334} {arXiv:0707.1334} \BibitemShut {NoStop}%
\bibitem [{\citenamefont {sesana}(2007)}]{sesana:2007zk}%
  \BibitemOpen
  \bibfield  {author} {\bibinfo {author} {\bibfnamefont {A.}~\bibnamefont
  {sesana}},\ }\href {\doibase 10.1111/j.1745-3933.2007.00375.x} {\bibfield
  {journal} {\bibinfo  {journal} {Mon. Not. Roy. Astron. Soc.}\ }\textbf
  {\bibinfo {volume} {382}},\ \bibinfo {pages} {6} (\bibinfo {year} {2007})},\
  \Eprint {http://arxiv.org/abs/0707.4677} {arXiv:0707.4677 [astro-ph]}
  \BibitemShut {NoStop}%
\bibitem [{\citenamefont {{Blecha}}\ and\ \citenamefont
  {{Loeb}}(2008)}]{Blecha:2008mg}%
  \BibitemOpen
  \bibfield  {author} {\bibinfo {author} {\bibfnamefont {L.}~\bibnamefont
  {{Blecha}}}\ and\ \bibinfo {author} {\bibfnamefont {A.}~\bibnamefont
  {{Loeb}}},\ }\href {\doibase 10.1111/j.1365-2966.2008.13790.x} {\bibfield
  {journal} {\bibinfo  {journal} {mnras}\ }\textbf {\bibinfo {volume} {390}},\
  \bibinfo {pages} {1311} (\bibinfo {year} {2008})},\ \Eprint
  {http://arxiv.org/abs/0805.1420} {arXiv:0805.1420} \BibitemShut {NoStop}%
\bibitem [{\citenamefont {Civano}\ \emph {et~al.}(2010)\citenamefont {Civano}
  \emph {et~al.}}]{Civano:2010es}%
  \BibitemOpen
  \bibfield  {author} {\bibinfo {author} {\bibfnamefont {F.}~\bibnamefont
  {Civano}} \emph {et~al.},\ }\href {\doibase 10.1088/0004-637X/717/1/209}
  {\bibfield  {journal} {\bibinfo  {journal} {Astrophys. J.}\ }\textbf
  {\bibinfo {volume} {717}},\ \bibinfo {pages} {209} (\bibinfo {year}
  {2010})},\ \Eprint {http://arxiv.org/abs/1003.0020} {arXiv:1003.0020
  [Unknown]} \BibitemShut {NoStop}%
\bibitem [{\citenamefont {Blecha}\ \emph {et~al.}(2012)\citenamefont {Blecha},
  \citenamefont {Civano}, \citenamefont {Elvis},\ and\ \citenamefont
  {Loeb}}]{Blecha:2012kx}%
  \BibitemOpen
  \bibfield  {author} {\bibinfo {author} {\bibfnamefont {L.}~\bibnamefont
  {Blecha}}, \bibinfo {author} {\bibfnamefont {F.}~\bibnamefont {Civano}},
  \bibinfo {author} {\bibfnamefont {M.}~\bibnamefont {Elvis}}, \ and\ \bibinfo
  {author} {\bibfnamefont {A.}~\bibnamefont {Loeb}},\ }\href@noop {} {\
  (\bibinfo {year} {2012})},\ \Eprint {http://arxiv.org/abs/1205.6202}
  {arXiv:1205.6202 [astro-ph.CO]} \BibitemShut {NoStop}%
\bibitem [{\citenamefont {Lanzuisi}\ \emph {et~al.}(2013)\citenamefont
  {Lanzuisi} \emph {et~al.}}]{Lanzuisi:2013fza}%
  \BibitemOpen
  \bibfield  {author} {\bibinfo {author} {\bibfnamefont {G.}~\bibnamefont
  {Lanzuisi}} \emph {et~al.},\ }\href {\doibase 10.1088/0004-637X/778/1/62}
  {\bibfield  {journal} {\bibinfo  {journal} {Astrophys. J.}\ }\textbf
  {\bibinfo {volume} {778}},\ \bibinfo {pages} {62} (\bibinfo {year} {2013})},\
  \Eprint {http://arxiv.org/abs/1310.1399} {arXiv:1310.1399 [astro-ph.HE]}
  \BibitemShut {NoStop}%
\bibitem [{\citenamefont {Bogdanovi\'c}\ \emph {et~al.}(2009)\citenamefont
  {Bogdanovi\'c}, \citenamefont {Eracleous},\ and\ \citenamefont
  {Sigurdsson}}]{Bogdanovic:2008uz}%
  \BibitemOpen
  \bibfield  {author} {\bibinfo {author} {\bibfnamefont {T.}~\bibnamefont
  {Bogdanovi\'c}}, \bibinfo {author} {\bibfnamefont {M.}~\bibnamefont
  {Eracleous}}, \ and\ \bibinfo {author} {\bibfnamefont {S.}~\bibnamefont
  {Sigurdsson}},\ }\href {\doibase 10.1088/0004-637X/697/1/288} {\bibfield
  {journal} {\bibinfo  {journal} {Astrophys. J.}\ }\textbf {\bibinfo {volume}
  {697}},\ \bibinfo {pages} {288} (\bibinfo {year} {2009})},\ \Eprint
  {http://arxiv.org/abs/0809.3262} {arXiv:0809.3262 [astro-ph]} \BibitemShut
  {NoStop}%
\bibitem [{\citenamefont {Komossa}\ \emph {et~al.}(2008)\citenamefont
  {Komossa}, \citenamefont {Zhou},\ and\ \citenamefont {Lu}}]{Komossa:2008qd}%
  \BibitemOpen
  \bibfield  {author} {\bibinfo {author} {\bibfnamefont {S.}~\bibnamefont
  {Komossa}}, \bibinfo {author} {\bibfnamefont {H.}~\bibnamefont {Zhou}}, \
  and\ \bibinfo {author} {\bibfnamefont {H.}~\bibnamefont {Lu}},\ }\href@noop
  {} {\bibfield  {journal} {\bibinfo  {journal} {Astrop. J. Letters}\ }\textbf
  {\bibinfo {volume} {678}},\ \bibinfo {pages} {L81} (\bibinfo {year}
  {2008})},\ \Eprint {http://arxiv.org/abs/0804.4585} {arXiv:0804.4585
  [astro-ph]} \BibitemShut {NoStop}%
\bibitem [{\citenamefont {Vivek}\ \emph {et~al.}(2009)\citenamefont {Vivek},
  \citenamefont {Srianand}, \citenamefont {Noterdaeme}, \citenamefont {Mohan},\
  and\ \citenamefont {Kuriakose}}]{Vivek:2009mm}%
  \BibitemOpen
  \bibfield  {author} {\bibinfo {author} {\bibfnamefont {M.}~\bibnamefont
  {Vivek}}, \bibinfo {author} {\bibfnamefont {R.}~\bibnamefont {Srianand}},
  \bibinfo {author} {\bibfnamefont {P.}~\bibnamefont {Noterdaeme}}, \bibinfo
  {author} {\bibfnamefont {V.}~\bibnamefont {Mohan}}, \ and\ \bibinfo {author}
  {\bibfnamefont {V.}~\bibnamefont {Kuriakose}},\ }\href@noop {} {\bibfield
  {journal} {\bibinfo  {journal} {mnras}\ }\textbf {\bibinfo {volume} {400}},\
  \bibinfo {pages} {L6} (\bibinfo {year} {2009})},\ \Eprint
  {http://arxiv.org/abs/arXiv:0909.0018} {arXiv:arXiv:0909.0018 [astro-ph.CO]}
  \BibitemShut {NoStop}%
\bibitem [{\citenamefont {Shields}\ \emph
  {et~al.}(2009{\natexlab{a}})\citenamefont {Shields}, \citenamefont
  {Bonning},\ and\ \citenamefont {Salviander}}]{Shields:2008kn}%
  \BibitemOpen
  \bibfield  {author} {\bibinfo {author} {\bibfnamefont {G.~A.}\ \bibnamefont
  {Shields}}, \bibinfo {author} {\bibfnamefont {E.~W.}\ \bibnamefont
  {Bonning}}, \ and\ \bibinfo {author} {\bibfnamefont {S.}~\bibnamefont
  {Salviander}},\ }\href {\doibase 10.1088/0004-637X/696/2/1367} {\bibfield
  {journal} {\bibinfo  {journal} {Astrophys. J.}\ }\textbf {\bibinfo {volume}
  {696}},\ \bibinfo {pages} {1367} (\bibinfo {year} {2009}{\natexlab{a}})},\
  \Eprint {http://arxiv.org/abs/0810.2563} {arXiv:0810.2563 [astro-ph]}
  \BibitemShut {NoStop}%
\bibitem [{\citenamefont {Decarli}\ \emph {et~al.}(2014)\citenamefont
  {Decarli}, \citenamefont {Dotti}, \citenamefont {Mazzucchelli}, \citenamefont
  {Montuori},\ and\ \citenamefont {Volonteri}}]{Decarli:2014wca}%
  \BibitemOpen
  \bibfield  {author} {\bibinfo {author} {\bibfnamefont {R.}~\bibnamefont
  {Decarli}}, \bibinfo {author} {\bibfnamefont {M.}~\bibnamefont {Dotti}},
  \bibinfo {author} {\bibfnamefont {C.}~\bibnamefont {Mazzucchelli}}, \bibinfo
  {author} {\bibfnamefont {C.}~\bibnamefont {Montuori}}, \ and\ \bibinfo
  {author} {\bibfnamefont {M.}~\bibnamefont {Volonteri}},\ }\href {\doibase
  10.1093/mnras/stu1810} {\bibfield  {journal} {\bibinfo  {journal} {Mon. Not.
  Roy. Astron. Soc.}\ }\textbf {\bibinfo {volume} {445}},\ \bibinfo {pages}
  {1558} (\bibinfo {year} {2014})},\ \Eprint {http://arxiv.org/abs/1409.1585}
  {arXiv:1409.1585 [astro-ph.GA]} \BibitemShut {NoStop}%
\bibitem [{\citenamefont {Jonker}\ \emph {et~al.}(2010)\citenamefont {Jonker},
  \citenamefont {Torres}, \citenamefont {Fabian}, \citenamefont {Heida},
  \citenamefont {Miniutti},\ and\ \citenamefont {Pooley}}]{Jonker:2010ip}%
  \BibitemOpen
  \bibfield  {author} {\bibinfo {author} {\bibfnamefont {P.~G.}\ \bibnamefont
  {Jonker}}, \bibinfo {author} {\bibfnamefont {M.~A.~P.}\ \bibnamefont
  {Torres}}, \bibinfo {author} {\bibfnamefont {A.~C.}\ \bibnamefont {Fabian}},
  \bibinfo {author} {\bibfnamefont {M.}~\bibnamefont {Heida}}, \bibinfo
  {author} {\bibfnamefont {G.}~\bibnamefont {Miniutti}}, \ and\ \bibinfo
  {author} {\bibfnamefont {D.}~\bibnamefont {Pooley}},\ }\href {\doibase
  10.1111/j.1365-2966.2010.16943.x} {\bibfield  {journal} {\bibinfo  {journal}
  {Mon. Not. Roy. Astron. Soc.}\ }\textbf {\bibinfo {volume} {407}},\ \bibinfo
  {pages} {645} (\bibinfo {year} {2010})},\ \Eprint
  {http://arxiv.org/abs/1004.5379} {arXiv:1004.5379 [astro-ph.HE]} \BibitemShut
  {NoStop}%
\bibitem [{\citenamefont {Shields}\ \emph
  {et~al.}(2009{\natexlab{b}})\citenamefont {Shields}, \citenamefont {Rosario},
  \citenamefont {Smith}, \citenamefont {Bonning}, \citenamefont {Salviander}
  \emph {et~al.}}]{Shields:2009jf}%
  \BibitemOpen
  \bibfield  {author} {\bibinfo {author} {\bibfnamefont {G.~A.}\ \bibnamefont
  {Shields}}, \bibinfo {author} {\bibfnamefont {D.~J.}\ \bibnamefont
  {Rosario}}, \bibinfo {author} {\bibfnamefont {K.~L.}\ \bibnamefont {Smith}},
  \bibinfo {author} {\bibfnamefont {E.~W.}\ \bibnamefont {Bonning}}, \bibinfo
  {author} {\bibfnamefont {S.}~\bibnamefont {Salviander}},  \emph {et~al.},\
  }\href {\doibase 10.1088/0004-637X/707/2/936} {\bibfield  {journal} {\bibinfo
   {journal} {Astrophys. J.}\ }\textbf {\bibinfo {volume} {707}},\ \bibinfo
  {pages} {936} (\bibinfo {year} {2009}{\natexlab{b}})},\ \Eprint
  {http://arxiv.org/abs/0907.3470} {arXiv:0907.3470 [astro-ph.CO]} \BibitemShut
  {NoStop}%
\bibitem [{\citenamefont {Shields}\ and\ \citenamefont
  {Bonning}(2013)}]{Shields:2013yaa}%
  \BibitemOpen
  \bibfield  {author} {\bibinfo {author} {\bibfnamefont {G.~A.}\ \bibnamefont
  {Shields}}\ and\ \bibinfo {author} {\bibfnamefont {E.~W.}\ \bibnamefont
  {Bonning}},\ }\href {\doibase 10.1088/2041-8205/772/1/L5} {\bibfield
  {journal} {\bibinfo  {journal} {Astrophys. J.}\ }\textbf {\bibinfo {volume}
  {772}},\ \bibinfo {pages} {L5} (\bibinfo {year} {2013})},\ \Eprint
  {http://arxiv.org/abs/1302.4458} {arXiv:1302.4458 [astro-ph.CO]} \BibitemShut
  {NoStop}%
\bibitem [{\citenamefont {Eracleous}\ \emph {et~al.}(2012)\citenamefont
  {Eracleous}, \citenamefont {Boroson}, \citenamefont {Halpern},\ and\
  \citenamefont {Liu}}]{Eracleous:2011ua}%
  \BibitemOpen
  \bibfield  {author} {\bibinfo {author} {\bibfnamefont {M.}~\bibnamefont
  {Eracleous}}, \bibinfo {author} {\bibfnamefont {T.~A.}\ \bibnamefont
  {Boroson}}, \bibinfo {author} {\bibfnamefont {J.~P.}\ \bibnamefont
  {Halpern}}, \ and\ \bibinfo {author} {\bibfnamefont {J.}~\bibnamefont
  {Liu}},\ }\href {\doibase 10.1088/0067-0049/201/2/23} {\bibfield  {journal}
  {\bibinfo  {journal} {The Astrophysical Journal Supplement}\ }\textbf
  {\bibinfo {volume} {201}},\ \bibinfo {eid} {23} (\bibinfo {year} {2012})},\
  \Eprint {http://arxiv.org/abs/1106.2952} {arXiv:1106.2952 [astro-ph.CO]}
  \BibitemShut {NoStop}%
\bibitem [{\citenamefont {Lena}\ \emph {et~al.}(2014)\citenamefont {Lena},
  \citenamefont {Robinson}, \citenamefont {Marconi}, \citenamefont {Axon},
  \citenamefont {Capetti}, \citenamefont {Merritt},\ and\ \citenamefont
  {Batcheldor}}]{Lena:2014zpa}%
  \BibitemOpen
  \bibfield  {author} {\bibinfo {author} {\bibfnamefont {D.}~\bibnamefont
  {Lena}}, \bibinfo {author} {\bibfnamefont {A.}~\bibnamefont {Robinson}},
  \bibinfo {author} {\bibfnamefont {A.}~\bibnamefont {Marconi}}, \bibinfo
  {author} {\bibfnamefont {D.~J.}\ \bibnamefont {Axon}}, \bibinfo {author}
  {\bibfnamefont {A.}~\bibnamefont {Capetti}}, \bibinfo {author} {\bibfnamefont
  {D.}~\bibnamefont {Merritt}}, \ and\ \bibinfo {author} {\bibfnamefont
  {D.}~\bibnamefont {Batcheldor}},\ }\href {\doibase
  10.1088/0004-637X/795/2/146} {\bibfield  {journal} {\bibinfo  {journal}
  {Astrophys. J.}\ }\textbf {\bibinfo {volume} {795}},\ \bibinfo {pages} {146}
  (\bibinfo {year} {2014})},\ \Eprint {http://arxiv.org/abs/1409.3976}
  {arXiv:1409.3976 [astro-ph.GA]} \BibitemShut {NoStop}%
\bibitem [{\citenamefont {Runnoe}\ \emph {et~al.}(2015)\citenamefont {Runnoe},
  \citenamefont {Eracleous}, \citenamefont {Mathes}, \citenamefont {Pennell},
  \citenamefont {Boroson}, \citenamefont {Sigurdsson}, \citenamefont
  {Bogdanovic}, \citenamefont {Halpern},\ and\ \citenamefont
  {Liu}}]{Runnoe:2015yba}%
  \BibitemOpen
  \bibfield  {author} {\bibinfo {author} {\bibfnamefont {J.~C.}\ \bibnamefont
  {Runnoe}}, \bibinfo {author} {\bibfnamefont {M.}~\bibnamefont {Eracleous}},
  \bibinfo {author} {\bibfnamefont {G.}~\bibnamefont {Mathes}}, \bibinfo
  {author} {\bibfnamefont {A.}~\bibnamefont {Pennell}}, \bibinfo {author}
  {\bibfnamefont {T.}~\bibnamefont {Boroson}}, \bibinfo {author} {\bibfnamefont
  {S.}~\bibnamefont {Sigurdsson}}, \bibinfo {author} {\bibfnamefont
  {T.}~\bibnamefont {Bogdanovic}}, \bibinfo {author} {\bibfnamefont {J.~P.}\
  \bibnamefont {Halpern}}, \ and\ \bibinfo {author} {\bibfnamefont
  {J.}~\bibnamefont {Liu}},\ }\href {\doibase 10.1088/0067-0049/221/1/7}
  {\bibfield  {journal} {\bibinfo  {journal} {Astrophys. J. Suppl.}\ }\textbf
  {\bibinfo {volume} {221}},\ \bibinfo {pages} {7} (\bibinfo {year} {2015})},\
  \Eprint {http://arxiv.org/abs/1509.02575} {arXiv:1509.02575 [astro-ph.GA]}
  \BibitemShut {NoStop}%
\bibitem [{\citenamefont {Runnoe}\ \emph {et~al.}(2017)\citenamefont {Runnoe},
  \citenamefont {Eracleous}, \citenamefont {Pennell}, \citenamefont {Mathes},
  \citenamefont {Boroson}, \citenamefont {Sigurdsson}, \citenamefont
  {Bogdanovi\'c}, \citenamefont {Halpern}, \citenamefont {Liu},\ and\
  \citenamefont {Brown}}]{Runnoe:2017oxn}%
  \BibitemOpen
  \bibfield  {author} {\bibinfo {author} {\bibfnamefont {J.~C.}\ \bibnamefont
  {Runnoe}}, \bibinfo {author} {\bibfnamefont {M.}~\bibnamefont {Eracleous}},
  \bibinfo {author} {\bibfnamefont {A.}~\bibnamefont {Pennell}}, \bibinfo
  {author} {\bibfnamefont {G.}~\bibnamefont {Mathes}}, \bibinfo {author}
  {\bibfnamefont {T.}~\bibnamefont {Boroson}}, \bibinfo {author} {\bibfnamefont
  {S.}~\bibnamefont {Sigurdsson}}, \bibinfo {author} {\bibfnamefont
  {T.}~\bibnamefont {Bogdanovi\'c}}, \bibinfo {author} {\bibfnamefont {J.~P.}\
  \bibnamefont {Halpern}}, \bibinfo {author} {\bibfnamefont {J.}~\bibnamefont
  {Liu}}, \ and\ \bibinfo {author} {\bibfnamefont {S.}~\bibnamefont {Brown}},\
  }\href {\doibase 10.1093/mnras/stx452} {\bibfield  {journal} {\bibinfo
  {journal} {Mon. Not. Roy. Astron. Soc.}\ }\textbf {\bibinfo {volume} {468}},\
  \bibinfo {pages} {1683} (\bibinfo {year} {2017})},\ \Eprint
  {http://arxiv.org/abs/1702.05465} {arXiv:1702.05465 [astro-ph.GA]}
  \BibitemShut {NoStop}%
\bibitem [{\citenamefont {Chiaberge}\ \emph {et~al.}(2017)\citenamefont
  {Chiaberge} \emph {et~al.}}]{Chiaberge:2016eqf}%
  \BibitemOpen
  \bibfield  {author} {\bibinfo {author} {\bibfnamefont {M.}~\bibnamefont
  {Chiaberge}} \emph {et~al.},\ }\href {\doibase 10.1051/0004-6361/201629522}
  {\bibfield  {journal} {\bibinfo  {journal} {Astron. Astrophys.}\ }\textbf
  {\bibinfo {volume} {600}},\ \bibinfo {pages} {A57} (\bibinfo {year}
  {2017})},\ \Eprint {http://arxiv.org/abs/1611.05501} {arXiv:1611.05501
  [astro-ph.GA]} \BibitemShut {NoStop}%
\bibitem [{\citenamefont {Lousto}\ \emph {et~al.}(2017)\citenamefont {Lousto},
  \citenamefont {Zlochower},\ and\ \citenamefont
  {Campanelli}}]{Lousto:2017uav}%
  \BibitemOpen
  \bibfield  {author} {\bibinfo {author} {\bibfnamefont {C.~O.}\ \bibnamefont
  {Lousto}}, \bibinfo {author} {\bibfnamefont {Y.}~\bibnamefont {Zlochower}}, \
  and\ \bibinfo {author} {\bibfnamefont {M.}~\bibnamefont {Campanelli}},\
  }\href {\doibase 10.3847/2041-8213/aa733c} {\bibfield  {journal} {\bibinfo
  {journal} {Astrophys. J.}\ }\textbf {\bibinfo {volume} {841}},\ \bibinfo
  {pages} {L28} (\bibinfo {year} {2017})},\ \Eprint
  {http://arxiv.org/abs/1704.00809} {arXiv:1704.00809 [astro-ph.GA]}
  \BibitemShut {NoStop}%
\bibitem [{\citenamefont {Chiaberge}\ \emph {et~al.}(2018)\citenamefont
  {Chiaberge}, \citenamefont {Tremblay}, \citenamefont {Capetti},\ and\
  \citenamefont {Norman}}]{Chiaberge:2018lkg}%
  \BibitemOpen
  \bibfield  {author} {\bibinfo {author} {\bibfnamefont {M.}~\bibnamefont
  {Chiaberge}}, \bibinfo {author} {\bibfnamefont {G.~R.}\ \bibnamefont
  {Tremblay}}, \bibinfo {author} {\bibfnamefont {A.}~\bibnamefont {Capetti}}, \
  and\ \bibinfo {author} {\bibfnamefont {C.}~\bibnamefont {Norman}},\ }\href
  {\doibase 10.3847/1538-4357/aac48b} {\bibfield  {journal} {\bibinfo
  {journal} {Astrophys. J.}\ }\textbf {\bibinfo {volume} {861}},\ \bibinfo
  {pages} {56} (\bibinfo {year} {2018})},\ \Eprint
  {http://arxiv.org/abs/1805.05860} {arXiv:1805.05860 [astro-ph.GA]}
  \BibitemShut {NoStop}%
\bibitem [{\citenamefont {Komossa}(2012)}]{Komossa:2012cy}%
  \BibitemOpen
  \bibfield  {author} {\bibinfo {author} {\bibfnamefont {S.}~\bibnamefont
  {Komossa}},\ }\href@noop {} {\bibfield  {journal} {\bibinfo  {journal} {Adv.
  Astron.}\ }\textbf {\bibinfo {volume} {2012}},\ \bibinfo {pages} {364973}
  (\bibinfo {year} {2012})},\ \Eprint {http://arxiv.org/abs/1202.1977}
  {arXiv:1202.1977 [astro-ph.CO]} \BibitemShut {NoStop}%
\bibitem [{\citenamefont {Blecha}\ \emph {et~al.}(2016)\citenamefont {Blecha},
  \citenamefont {Sijacki}, \citenamefont {Kelley}, \citenamefont {Torrey},
  \citenamefont {Vogelsberger}, \citenamefont {Nelson}, \citenamefont
  {Springel}, \citenamefont {Snyder},\ and\ \citenamefont
  {Hernquist}}]{Blecha:2015baa}%
  \BibitemOpen
  \bibfield  {author} {\bibinfo {author} {\bibfnamefont {L.}~\bibnamefont
  {Blecha}}, \bibinfo {author} {\bibfnamefont {D.}~\bibnamefont {Sijacki}},
  \bibinfo {author} {\bibfnamefont {L.~Z.}\ \bibnamefont {Kelley}}, \bibinfo
  {author} {\bibfnamefont {P.}~\bibnamefont {Torrey}}, \bibinfo {author}
  {\bibfnamefont {M.}~\bibnamefont {Vogelsberger}}, \bibinfo {author}
  {\bibfnamefont {D.}~\bibnamefont {Nelson}}, \bibinfo {author} {\bibfnamefont
  {V.}~\bibnamefont {Springel}}, \bibinfo {author} {\bibfnamefont
  {G.}~\bibnamefont {Snyder}}, \ and\ \bibinfo {author} {\bibfnamefont
  {L.}~\bibnamefont {Hernquist}},\ }\href {\doibase 10.1093/mnras/stv2646}
  {\bibfield  {journal} {\bibinfo  {journal} {Mon. Not. Roy. Astron. Soc.}\
  }\textbf {\bibinfo {volume} {456}},\ \bibinfo {pages} {961} (\bibinfo {year}
  {2016})},\ \Eprint {http://arxiv.org/abs/1508.01524} {arXiv:1508.01524
  [astro-ph.GA]} \BibitemShut {NoStop}%
\bibitem [{\citenamefont {Lousto}\ \emph {et~al.}(2012)\citenamefont {Lousto},
  \citenamefont {Zlochower}, \citenamefont {Dotti},\ and\ \citenamefont
  {Volonteri}}]{Lousto:2012su}%
  \BibitemOpen
  \bibfield  {author} {\bibinfo {author} {\bibfnamefont {C.~O.}\ \bibnamefont
  {Lousto}}, \bibinfo {author} {\bibfnamefont {Y.}~\bibnamefont {Zlochower}},
  \bibinfo {author} {\bibfnamefont {M.}~\bibnamefont {Dotti}}, \ and\ \bibinfo
  {author} {\bibfnamefont {M.}~\bibnamefont {Volonteri}},\ }\href@noop {}
  {\bibfield  {journal} {\bibinfo  {journal} {Phys. Rev.}\ }\textbf {\bibinfo
  {volume} {D85}},\ \bibinfo {pages} {084015} (\bibinfo {year} {2012})},\
  \Eprint {http://arxiv.org/abs/1201.1923} {arXiv:1201.1923 [gr-qc]}
  \BibitemShut {NoStop}%
\bibitem [{\citenamefont {Zlochower}\ and\ \citenamefont
  {Lousto}(2015)}]{Zlochower:2015wga}%
  \BibitemOpen
  \bibfield  {author} {\bibinfo {author} {\bibfnamefont {Y.}~\bibnamefont
  {Zlochower}}\ and\ \bibinfo {author} {\bibfnamefont {C.~O.}\ \bibnamefont
  {Lousto}},\ }\href {\doibase 10.1103/PhysRevD.92.024022} {\bibfield
  {journal} {\bibinfo  {journal} {Phys. Rev.}\ }\textbf {\bibinfo {volume}
  {D92}},\ \bibinfo {pages} {024022} (\bibinfo {year} {2015})},\ \Eprint
  {http://arxiv.org/abs/1503.07536} {arXiv:1503.07536 [gr-qc]} \BibitemShut
  {NoStop}%
\bibitem [{\citenamefont {Lousto}\ and\ \citenamefont
  {Zlochower}(2011{\natexlab{a}})}]{Lousto:2011kp}%
  \BibitemOpen
  \bibfield  {author} {\bibinfo {author} {\bibfnamefont {C.~O.}\ \bibnamefont
  {Lousto}}\ and\ \bibinfo {author} {\bibfnamefont {Y.}~\bibnamefont
  {Zlochower}},\ }\href {\doibase 10.1103/PhysRevLett.107.231102} {\bibfield
  {journal} {\bibinfo  {journal} {Phys. Rev. Lett.}\ }\textbf {\bibinfo
  {volume} {107}},\ \bibinfo {pages} {231102} (\bibinfo {year}
  {2011}{\natexlab{a}})},\ \Eprint {http://arxiv.org/abs/1108.2009}
  {arXiv:1108.2009 [gr-qc]} \BibitemShut {NoStop}%
\bibitem [{\citenamefont {Healy}\ \emph {et~al.}(2014)\citenamefont {Healy},
  \citenamefont {Lousto},\ and\ \citenamefont {Zlochower}}]{Healy:2014yta}%
  \BibitemOpen
  \bibfield  {author} {\bibinfo {author} {\bibfnamefont {J.}~\bibnamefont
  {Healy}}, \bibinfo {author} {\bibfnamefont {C.~O.}\ \bibnamefont {Lousto}}, \
  and\ \bibinfo {author} {\bibfnamefont {Y.}~\bibnamefont {Zlochower}},\ }\href
  {\doibase 10.1103/PhysRevD.90.104004} {\bibfield  {journal} {\bibinfo
  {journal} {Phys. Rev.}\ }\textbf {\bibinfo {volume} {D90}},\ \bibinfo {pages}
  {104004} (\bibinfo {year} {2014})},\ \Eprint {http://arxiv.org/abs/1406.7295}
  {arXiv:1406.7295 [gr-qc]} \BibitemShut {NoStop}%
\bibitem [{\citenamefont {Healy}\ \emph
  {et~al.}(2018{\natexlab{a}})\citenamefont {Healy}, \citenamefont {Lousto},
  \citenamefont {Ruchlin},\ and\ \citenamefont {Zlochower}}]{Healy:2017vuz}%
  \BibitemOpen
  \bibfield  {author} {\bibinfo {author} {\bibfnamefont {J.}~\bibnamefont
  {Healy}}, \bibinfo {author} {\bibfnamefont {C.~O.}\ \bibnamefont {Lousto}},
  \bibinfo {author} {\bibfnamefont {I.}~\bibnamefont {Ruchlin}}, \ and\
  \bibinfo {author} {\bibfnamefont {Y.}~\bibnamefont {Zlochower}},\ }\href
  {\doibase 10.1103/PhysRevD.97.104026} {\bibfield  {journal} {\bibinfo
  {journal} {Phys. Rev.}\ }\textbf {\bibinfo {volume} {D97}},\ \bibinfo {pages}
  {104026} (\bibinfo {year} {2018}{\natexlab{a}})},\ \Eprint
  {http://arxiv.org/abs/1711.09041} {arXiv:1711.09041 [gr-qc]} \BibitemShut
  {NoStop}%
\bibitem [{\citenamefont {Gonz\'alez}\ \emph {et~al.}(2007)\citenamefont
  {Gonz\'alez}, \citenamefont {Sperhake}, \citenamefont {Brugmann},
  \citenamefont {Hannam},\ and\ \citenamefont {Husa}}]{Gonzalez:2006md}%
  \BibitemOpen
  \bibfield  {author} {\bibinfo {author} {\bibfnamefont {J.~A.}\ \bibnamefont
  {Gonz\'alez}}, \bibinfo {author} {\bibfnamefont {U.}~\bibnamefont
  {Sperhake}}, \bibinfo {author} {\bibfnamefont {B.}~\bibnamefont {Brugmann}},
  \bibinfo {author} {\bibfnamefont {M.}~\bibnamefont {Hannam}}, \ and\ \bibinfo
  {author} {\bibfnamefont {S.}~\bibnamefont {Husa}},\ }\href@noop {} {\bibfield
   {journal} {\bibinfo  {journal} {Phys. Rev. Lett.}\ }\textbf {\bibinfo
  {volume} {98}},\ \bibinfo {pages} {091101} (\bibinfo {year} {2007})},\
  \Eprint {http://arxiv.org/abs/gr-qc/0610154} {gr-qc/0610154} \BibitemShut
  {NoStop}%
\bibitem [{\citenamefont {Healy}\ \emph
  {et~al.}(2017{\natexlab{a}})\citenamefont {Healy}, \citenamefont {Lousto},\
  and\ \citenamefont {Zlochower}}]{Healy:2017mvh}%
  \BibitemOpen
  \bibfield  {author} {\bibinfo {author} {\bibfnamefont {J.}~\bibnamefont
  {Healy}}, \bibinfo {author} {\bibfnamefont {C.~O.}\ \bibnamefont {Lousto}}, \
  and\ \bibinfo {author} {\bibfnamefont {Y.}~\bibnamefont {Zlochower}},\ }\href
  {\doibase 10.1103/PhysRevD.96.024031} {\bibfield  {journal} {\bibinfo
  {journal} {Phys. Rev.}\ }\textbf {\bibinfo {volume} {D96}},\ \bibinfo {pages}
  {024031} (\bibinfo {year} {2017}{\natexlab{a}})},\ \Eprint
  {http://arxiv.org/abs/1705.07034} {arXiv:1705.07034 [gr-qc]} \BibitemShut
  {NoStop}%
\bibitem [{\citenamefont {Sperhake}(2015)}]{Sperhake:2014wpa}%
  \BibitemOpen
  \bibfield  {author} {\bibinfo {author} {\bibfnamefont {U.}~\bibnamefont
  {Sperhake}},\ }\href {\doibase 10.1088/0264-9381/32/12/124011} {\bibfield
  {journal} {\bibinfo  {journal} {Class. Quant. Grav.}\ }\textbf {\bibinfo
  {volume} {32}},\ \bibinfo {pages} {124011} (\bibinfo {year} {2015})},\
  \Eprint {http://arxiv.org/abs/1411.3997} {arXiv:1411.3997 [gr-qc]}
  \BibitemShut {NoStop}%
\bibitem [{\citenamefont {Sijacki}\ \emph {et~al.}(2011)\citenamefont
  {Sijacki}, \citenamefont {Springel},\ and\ \citenamefont
  {Haehnelt}}]{Sijacki:2010tk}%
  \BibitemOpen
  \bibfield  {author} {\bibinfo {author} {\bibfnamefont {D.}~\bibnamefont
  {Sijacki}}, \bibinfo {author} {\bibfnamefont {V.}~\bibnamefont {Springel}}, \
  and\ \bibinfo {author} {\bibfnamefont {M.}~\bibnamefont {Haehnelt}},\ }\href
  {\doibase 10.1111/j.1365-2966.2011.18666.x} {\bibfield  {journal} {\bibinfo
  {journal} {Mon. Not. Roy. Astron. Soc.}\ }\textbf {\bibinfo {volume} {414}},\
  \bibinfo {pages} {3656} (\bibinfo {year} {2011})},\ \Eprint
  {http://arxiv.org/abs/1008.3313} {arXiv:1008.3313 [astro-ph.CO]} \BibitemShut
  {NoStop}%
\bibitem [{\citenamefont {Ponce}\ \emph {et~al.}(2012)\citenamefont {Ponce},
  \citenamefont {Faber},\ and\ \citenamefont {Lombardi}}]{Ponce:2011kv}%
  \BibitemOpen
  \bibfield  {author} {\bibinfo {author} {\bibfnamefont {M.}~\bibnamefont
  {Ponce}}, \bibinfo {author} {\bibfnamefont {J.~A.}\ \bibnamefont {Faber}}, \
  and\ \bibinfo {author} {\bibfnamefont {J.}~\bibnamefont {Lombardi},
  \bibfnamefont {James~C.}},\ }\href {\doibase 10.1088/0004-637X/745/1/71}
  {\bibfield  {journal} {\bibinfo  {journal} {Astrophys. J.}\ }\textbf
  {\bibinfo {volume} {745}},\ \bibinfo {pages} {71} (\bibinfo {year} {2012})},\
  \Eprint {http://arxiv.org/abs/1107.1711} {arXiv:1107.1711 [astro-ph.CO]}
  \BibitemShut {NoStop}%
\bibitem [{\citenamefont {Meliani}\ \emph {et~al.}(2017)\citenamefont
  {Meliani}, \citenamefont {Mizuno}, \citenamefont {Olivares}, \citenamefont
  {Porth}, \citenamefont {Rezzolla},\ and\ \citenamefont
  {Younsi}}]{Meliani:2016rwn}%
  \BibitemOpen
  \bibfield  {author} {\bibinfo {author} {\bibfnamefont {Z.}~\bibnamefont
  {Meliani}}, \bibinfo {author} {\bibfnamefont {Y.}~\bibnamefont {Mizuno}},
  \bibinfo {author} {\bibfnamefont {H.}~\bibnamefont {Olivares}}, \bibinfo
  {author} {\bibfnamefont {O.}~\bibnamefont {Porth}}, \bibinfo {author}
  {\bibfnamefont {L.}~\bibnamefont {Rezzolla}}, \ and\ \bibinfo {author}
  {\bibfnamefont {Z.}~\bibnamefont {Younsi}},\ }\href {\doibase
  10.1051/0004-6361/201629191} {\bibfield  {journal} {\bibinfo  {journal}
  {Astron. Astrophys.}\ }\textbf {\bibinfo {volume} {598}},\ \bibinfo {pages}
  {A38} (\bibinfo {year} {2017})},\ \Eprint {http://arxiv.org/abs/1606.08192}
  {arXiv:1606.08192 [astro-ph.HE]} \BibitemShut {NoStop}%
\bibitem [{\citenamefont {Gerosa}\ and\ \citenamefont
  {Moore}(2016)}]{Gerosa:2016vip}%
  \BibitemOpen
  \bibfield  {author} {\bibinfo {author} {\bibfnamefont {D.}~\bibnamefont
  {Gerosa}}\ and\ \bibinfo {author} {\bibfnamefont {C.~J.}\ \bibnamefont
  {Moore}},\ }\href {\doibase 10.1103/PhysRevLett.117.011101} {\bibfield
  {journal} {\bibinfo  {journal} {Phys. Rev. Lett.}\ }\textbf {\bibinfo
  {volume} {117}},\ \bibinfo {pages} {011101} (\bibinfo {year} {2016})},\
  \Eprint {http://arxiv.org/abs/1606.04226} {arXiv:1606.04226 [gr-qc]}
  \BibitemShut {NoStop}%
\bibitem [{\citenamefont {Calder\'on~Bustillo}\ \emph
  {et~al.}(2018)\citenamefont {Calder\'on~Bustillo}, \citenamefont {Clark},
  \citenamefont {Laguna},\ and\ \citenamefont
  {Shoemaker}}]{CalderonBustillo:2018zuq}%
  \BibitemOpen
  \bibfield  {author} {\bibinfo {author} {\bibfnamefont {J.}~\bibnamefont
  {Calder\'on~Bustillo}}, \bibinfo {author} {\bibfnamefont {J.~A.}\
  \bibnamefont {Clark}}, \bibinfo {author} {\bibfnamefont {P.}~\bibnamefont
  {Laguna}}, \ and\ \bibinfo {author} {\bibfnamefont {D.}~\bibnamefont
  {Shoemaker}},\ }\href {\doibase 10.1103/PhysRevLett.121.191102} {\bibfield
  {journal} {\bibinfo  {journal} {Phys. Rev. Lett.}\ }\textbf {\bibinfo
  {volume} {121}},\ \bibinfo {pages} {191102} (\bibinfo {year} {2018})},\
  \Eprint {http://arxiv.org/abs/1806.11160} {arXiv:1806.11160 [gr-qc]}
  \BibitemShut {NoStop}%
\bibitem [{Adv()}]{AdvLIGONC}%
  \BibitemOpen
  \href {https://dcc.ligo.org/cgi-bin/DocDB/ShowDocument?docid=2974} {\enquote
  {\bibinfo {title} {Advanced ligo anticipated sensitivity curves},}\
  }\BibitemShut {NoStop}%
\bibitem [{\citenamefont {Healy}\ \emph {et~al.}(2019)\citenamefont {Healy},
  \citenamefont {Lousto}, \citenamefont {Lange}, \citenamefont {O'Shaughnessy},
  \citenamefont {Zlochower},\ and\ \citenamefont {Campanelli}}]{Healy:2019jyf}%
  \BibitemOpen
  \bibfield  {author} {\bibinfo {author} {\bibfnamefont {J.}~\bibnamefont
  {Healy}}, \bibinfo {author} {\bibfnamefont {C.~O.}\ \bibnamefont {Lousto}},
  \bibinfo {author} {\bibfnamefont {J.}~\bibnamefont {Lange}}, \bibinfo
  {author} {\bibfnamefont {R.}~\bibnamefont {O'Shaughnessy}}, \bibinfo {author}
  {\bibfnamefont {Y.}~\bibnamefont {Zlochower}}, \ and\ \bibinfo {author}
  {\bibfnamefont {M.}~\bibnamefont {Campanelli}},\ }\href {\doibase
  10.1103/PhysRevD.100.024021} {\bibfield  {journal} {\bibinfo  {journal}
  {Phys. Rev.}\ }\textbf {\bibinfo {volume} {D100}},\ \bibinfo {pages} {024021}
  (\bibinfo {year} {2019})},\ \Eprint {http://arxiv.org/abs/1901.02553}
  {arXiv:1901.02553 [gr-qc]} \BibitemShut {NoStop}%
\bibitem [{\citenamefont {Pretorius}(2005)}]{Pretorius:2005gq}%
  \BibitemOpen
  \bibfield  {author} {\bibinfo {author} {\bibfnamefont {F.}~\bibnamefont
  {Pretorius}},\ }\href@noop {} {\bibfield  {journal} {\bibinfo  {journal}
  {Phys. Rev. Lett.}\ }\textbf {\bibinfo {volume} {95}},\ \bibinfo {pages}
  {121101} (\bibinfo {year} {2005})},\ \Eprint
  {http://arxiv.org/abs/gr-qc/0507014} {gr-qc/0507014} \BibitemShut {NoStop}%
\bibitem [{\citenamefont {Campanelli}\ \emph
  {et~al.}(2006{\natexlab{a}})\citenamefont {Campanelli}, \citenamefont
  {Lousto}, \citenamefont {Marronetti},\ and\ \citenamefont
  {Zlochower}}]{Campanelli:2005dd}%
  \BibitemOpen
  \bibfield  {author} {\bibinfo {author} {\bibfnamefont {M.}~\bibnamefont
  {Campanelli}}, \bibinfo {author} {\bibfnamefont {C.~O.}\ \bibnamefont
  {Lousto}}, \bibinfo {author} {\bibfnamefont {P.}~\bibnamefont {Marronetti}},
  \ and\ \bibinfo {author} {\bibfnamefont {Y.}~\bibnamefont {Zlochower}},\
  }\href@noop {} {\bibfield  {journal} {\bibinfo  {journal} {Phys. Rev. Lett.}\
  }\textbf {\bibinfo {volume} {96}},\ \bibinfo {pages} {111101} (\bibinfo
  {year} {2006}{\natexlab{a}})},\ \Eprint {http://arxiv.org/abs/gr-qc/0511048}
  {gr-qc/0511048} \BibitemShut {NoStop}%
\bibitem [{\citenamefont {Baker}\ \emph {et~al.}(2006)\citenamefont {Baker},
  \citenamefont {Centrella}, \citenamefont {Choi}, \citenamefont {Koppitz},\
  and\ \citenamefont {van Meter}}]{Baker:2005vv}%
  \BibitemOpen
  \bibfield  {author} {\bibinfo {author} {\bibfnamefont {J.~G.}\ \bibnamefont
  {Baker}}, \bibinfo {author} {\bibfnamefont {J.}~\bibnamefont {Centrella}},
  \bibinfo {author} {\bibfnamefont {D.-I.}\ \bibnamefont {Choi}}, \bibinfo
  {author} {\bibfnamefont {M.}~\bibnamefont {Koppitz}}, \ and\ \bibinfo
  {author} {\bibfnamefont {J.}~\bibnamefont {van Meter}},\ }\href@noop {}
  {\bibfield  {journal} {\bibinfo  {journal} {Phys. Rev. Lett.}\ }\textbf
  {\bibinfo {volume} {96}},\ \bibinfo {pages} {111102} (\bibinfo {year}
  {2006})},\ \Eprint {http://arxiv.org/abs/gr-qc/0511103} {gr-qc/0511103}
  \BibitemShut {NoStop}%
\bibitem [{\citenamefont {Campanelli}\ \emph
  {et~al.}(2006{\natexlab{b}})\citenamefont {Campanelli}, \citenamefont
  {Lousto},\ and\ \citenamefont {Zlochower}}]{Campanelli:2006uy}%
  \BibitemOpen
  \bibfield  {author} {\bibinfo {author} {\bibfnamefont {M.}~\bibnamefont
  {Campanelli}}, \bibinfo {author} {\bibfnamefont {C.~O.}\ \bibnamefont
  {Lousto}}, \ and\ \bibinfo {author} {\bibfnamefont {Y.}~\bibnamefont
  {Zlochower}},\ }\href@noop {} {\bibfield  {journal} {\bibinfo  {journal}
  {Phys. Rev.}\ }\textbf {\bibinfo {volume} {D74}},\ \bibinfo {pages}
  {041501(R)} (\bibinfo {year} {2006}{\natexlab{b}})},\ \Eprint
  {http://arxiv.org/abs/gr-qc/0604012} {gr-qc/0604012} \BibitemShut {NoStop}%
\bibitem [{\citenamefont {Lousto}\ \emph {et~al.}(2016)\citenamefont {Lousto},
  \citenamefont {Healy},\ and\ \citenamefont {Nakano}}]{Lousto:2015uwa}%
  \BibitemOpen
  \bibfield  {author} {\bibinfo {author} {\bibfnamefont {C.~O.}\ \bibnamefont
  {Lousto}}, \bibinfo {author} {\bibfnamefont {J.}~\bibnamefont {Healy}}, \
  and\ \bibinfo {author} {\bibfnamefont {H.}~\bibnamefont {Nakano}},\ }\href
  {\doibase 10.1103/PhysRevD.93.044031} {\bibfield  {journal} {\bibinfo
  {journal} {Phys. Rev.}\ }\textbf {\bibinfo {volume} {D93}},\ \bibinfo {pages}
  {044031} (\bibinfo {year} {2016})},\ \Eprint
  {http://arxiv.org/abs/1506.04768} {arXiv:1506.04768 [gr-qc]} \BibitemShut
  {NoStop}%
\bibitem [{\citenamefont {Kesden}\ \emph {et~al.}(2015)\citenamefont {Kesden},
  \citenamefont {Gerosa}, \citenamefont {O'Shaughnessy}, \citenamefont
  {Berti},\ and\ \citenamefont {Sperhake}}]{Kesden:2014sla}%
  \BibitemOpen
  \bibfield  {author} {\bibinfo {author} {\bibfnamefont {M.}~\bibnamefont
  {Kesden}}, \bibinfo {author} {\bibfnamefont {D.}~\bibnamefont {Gerosa}},
  \bibinfo {author} {\bibfnamefont {R.}~\bibnamefont {O'Shaughnessy}}, \bibinfo
  {author} {\bibfnamefont {E.}~\bibnamefont {Berti}}, \ and\ \bibinfo {author}
  {\bibfnamefont {U.}~\bibnamefont {Sperhake}},\ }\href {\doibase
  10.1103/PhysRevLett.114.081103} {\bibfield  {journal} {\bibinfo  {journal}
  {Phys. Rev. Lett.}\ }\textbf {\bibinfo {volume} {114}},\ \bibinfo {pages}
  {081103} (\bibinfo {year} {2015})},\ \Eprint {http://arxiv.org/abs/1411.0674}
  {arXiv:1411.0674 [gr-qc]} \BibitemShut {NoStop}%
\bibitem [{\citenamefont {Lousto}\ and\ \citenamefont
  {Healy}(2016)}]{Lousto:2016nlp}%
  \BibitemOpen
  \bibfield  {author} {\bibinfo {author} {\bibfnamefont {C.~O.}\ \bibnamefont
  {Lousto}}\ and\ \bibinfo {author} {\bibfnamefont {J.}~\bibnamefont {Healy}},\
  }\href {\doibase 10.1103/PhysRevD.93.124074} {\bibfield  {journal} {\bibinfo
  {journal} {Phys. Rev.}\ }\textbf {\bibinfo {volume} {D93}},\ \bibinfo {pages}
  {124074} (\bibinfo {year} {2016})},\ \Eprint
  {http://arxiv.org/abs/1601.05086} {arXiv:1601.05086 [gr-qc]} \BibitemShut
  {NoStop}%
\bibitem [{\citenamefont {Lousto}\ and\ \citenamefont
  {Healy}(2019)}]{Lousto:2018dgd}%
  \BibitemOpen
  \bibfield  {author} {\bibinfo {author} {\bibfnamefont {C.~O.}\ \bibnamefont
  {Lousto}}\ and\ \bibinfo {author} {\bibfnamefont {J.}~\bibnamefont {Healy}},\
  }\href {\doibase 10.1103/PhysRevD.99.064023} {\bibfield  {journal} {\bibinfo
  {journal} {Phys. Rev.}\ }\textbf {\bibinfo {volume} {D99}},\ \bibinfo {pages}
  {064023} (\bibinfo {year} {2019})},\ \Eprint
  {http://arxiv.org/abs/1805.08127} {arXiv:1805.08127 [gr-qc]} \BibitemShut
  {NoStop}%
\bibitem [{\citenamefont {Ruchlin}\ \emph {et~al.}(2017)\citenamefont
  {Ruchlin}, \citenamefont {Healy}, \citenamefont {Lousto},\ and\ \citenamefont
  {Zlochower}}]{Ruchlin:2014zva}%
  \BibitemOpen
  \bibfield  {author} {\bibinfo {author} {\bibfnamefont {I.}~\bibnamefont
  {Ruchlin}}, \bibinfo {author} {\bibfnamefont {J.}~\bibnamefont {Healy}},
  \bibinfo {author} {\bibfnamefont {C.~O.}\ \bibnamefont {Lousto}}, \ and\
  \bibinfo {author} {\bibfnamefont {Y.}~\bibnamefont {Zlochower}},\ }\href
  {\doibase 10.1103/PhysRevD.95.024033} {\bibfield  {journal} {\bibinfo
  {journal} {Phys. Rev.}\ }\textbf {\bibinfo {volume} {D95}},\ \bibinfo {pages}
  {024033} (\bibinfo {year} {2017})},\ \Eprint {http://arxiv.org/abs/1410.8607}
  {arXiv:1410.8607 [gr-qc]} \BibitemShut {NoStop}%
\bibitem [{\citenamefont {Healy}\ \emph {et~al.}(2016)\citenamefont {Healy},
  \citenamefont {Ruchlin}, \citenamefont {Lousto},\ and\ \citenamefont
  {Zlochower}}]{Healy:2015mla}%
  \BibitemOpen
  \bibfield  {author} {\bibinfo {author} {\bibfnamefont {J.}~\bibnamefont
  {Healy}}, \bibinfo {author} {\bibfnamefont {I.}~\bibnamefont {Ruchlin}},
  \bibinfo {author} {\bibfnamefont {C.~O.}\ \bibnamefont {Lousto}}, \ and\
  \bibinfo {author} {\bibfnamefont {Y.}~\bibnamefont {Zlochower}},\ }\href
  {\doibase 10.1103/PhysRevD.94.104020} {\bibfield  {journal} {\bibinfo
  {journal} {Phys. Rev.}\ }\textbf {\bibinfo {volume} {D94}},\ \bibinfo {pages}
  {104020} (\bibinfo {year} {2016})},\ \Eprint
  {http://arxiv.org/abs/1506.06153} {arXiv:1506.06153 [gr-qc]} \BibitemShut
  {NoStop}%
\bibitem [{\citenamefont {Zlochower}\ \emph {et~al.}(2017)\citenamefont
  {Zlochower}, \citenamefont {Healy}, \citenamefont {Lousto},\ and\
  \citenamefont {Ruchlin}}]{Zlochower:2017bbg}%
  \BibitemOpen
  \bibfield  {author} {\bibinfo {author} {\bibfnamefont {Y.}~\bibnamefont
  {Zlochower}}, \bibinfo {author} {\bibfnamefont {J.}~\bibnamefont {Healy}},
  \bibinfo {author} {\bibfnamefont {C.~O.}\ \bibnamefont {Lousto}}, \ and\
  \bibinfo {author} {\bibfnamefont {I.}~\bibnamefont {Ruchlin}},\ }\href
  {\doibase 10.1103/PhysRevD.96.044002} {\bibfield  {journal} {\bibinfo
  {journal} {Phys. Rev.}\ }\textbf {\bibinfo {volume} {D96}},\ \bibinfo {pages}
  {044002} (\bibinfo {year} {2017})},\ \Eprint
  {http://arxiv.org/abs/1706.01980} {arXiv:1706.01980 [gr-qc]} \BibitemShut
  {NoStop}%
\bibitem [{\citenamefont {Mroue}\ \emph {et~al.}(2013)\citenamefont {Mroue},
  \citenamefont {Scheel}, \citenamefont {Szilagyi}, \citenamefont {Pfeiffer},
  \citenamefont {Boyle} \emph {et~al.}}]{Mroue:2013xna}%
  \BibitemOpen
  \bibfield  {author} {\bibinfo {author} {\bibfnamefont {A.~H.}\ \bibnamefont
  {Mroue}}, \bibinfo {author} {\bibfnamefont {M.~A.}\ \bibnamefont {Scheel}},
  \bibinfo {author} {\bibfnamefont {B.}~\bibnamefont {Szilagyi}}, \bibinfo
  {author} {\bibfnamefont {H.~P.}\ \bibnamefont {Pfeiffer}}, \bibinfo {author}
  {\bibfnamefont {M.}~\bibnamefont {Boyle}},  \emph {et~al.},\ }\href {\doibase
  10.1103/PhysRevLett.111.241104} {\bibfield  {journal} {\bibinfo  {journal}
  {Phys. Rev. Lett.}\ }\textbf {\bibinfo {volume} {111}},\ \bibinfo {pages}
  {241104} (\bibinfo {year} {2013})},\ \Eprint {http://arxiv.org/abs/1304.6077}
  {arXiv:1304.6077 [gr-qc]} \BibitemShut {NoStop}%
\bibitem [{\citenamefont {Zlochower}\ \emph {et~al.}(2005)\citenamefont
  {Zlochower}, \citenamefont {Baker}, \citenamefont {Campanelli},\ and\
  \citenamefont {Lousto}}]{Zlochower:2005bj}%
  \BibitemOpen
  \bibfield  {author} {\bibinfo {author} {\bibfnamefont {Y.}~\bibnamefont
  {Zlochower}}, \bibinfo {author} {\bibfnamefont {J.~G.}\ \bibnamefont
  {Baker}}, \bibinfo {author} {\bibfnamefont {M.}~\bibnamefont {Campanelli}}, \
  and\ \bibinfo {author} {\bibfnamefont {C.~O.}\ \bibnamefont {Lousto}},\
  }\href {\doibase 10.1103/PhysRevD.72.024021} {\bibfield  {journal} {\bibinfo
  {journal} {Phys. Rev.}\ }\textbf {\bibinfo {volume} {D72}},\ \bibinfo {pages}
  {024021} (\bibinfo {year} {2005})},\ \Eprint
  {http://arxiv.org/abs/gr-qc/0505055} {arXiv:gr-qc/0505055} \BibitemShut
  {NoStop}%
\bibitem [{\citenamefont {Marronetti}\ \emph {et~al.}(2008)\citenamefont
  {Marronetti}, \citenamefont {Tichy}, \citenamefont {Br{\"u}gmann},
  \citenamefont {Gonzalez},\ and\ \citenamefont
  {Sperhake}}]{Marronetti:2007wz}%
  \BibitemOpen
  \bibfield  {author} {\bibinfo {author} {\bibfnamefont {P.}~\bibnamefont
  {Marronetti}}, \bibinfo {author} {\bibfnamefont {W.}~\bibnamefont {Tichy}},
  \bibinfo {author} {\bibfnamefont {B.}~\bibnamefont {Br{\"u}gmann}}, \bibinfo
  {author} {\bibfnamefont {J.}~\bibnamefont {Gonzalez}}, \ and\ \bibinfo
  {author} {\bibfnamefont {U.}~\bibnamefont {Sperhake}},\ }\href {\doibase
  10.1103/PhysRevD.77.064010} {\bibfield  {journal} {\bibinfo  {journal} {Phys.
  Rev.}\ }\textbf {\bibinfo {volume} {D77}},\ \bibinfo {pages} {064010}
  (\bibinfo {year} {2008})},\ \Eprint {http://arxiv.org/abs/0709.2160}
  {arXiv:0709.2160 [gr-qc]} \BibitemShut {NoStop}%
\bibitem [{\citenamefont {Lousto}\ and\ \citenamefont
  {Zlochower}(2008)}]{Lousto:2007rj}%
  \BibitemOpen
  \bibfield  {author} {\bibinfo {author} {\bibfnamefont {C.~O.}\ \bibnamefont
  {Lousto}}\ and\ \bibinfo {author} {\bibfnamefont {Y.}~\bibnamefont
  {Zlochower}},\ }\href {\doibase 10.1103/PhysRevD.77.024034} {\bibfield
  {journal} {\bibinfo  {journal} {Phys. Rev.}\ }\textbf {\bibinfo {volume}
  {D77}},\ \bibinfo {pages} {024034} (\bibinfo {year} {2008})},\ \Eprint
  {http://arxiv.org/abs/0711.1165} {arXiv:0711.1165 [gr-qc]} \BibitemShut
  {NoStop}%
\bibitem [{\citenamefont {Alic}\ \emph {et~al.}(2012)\citenamefont {Alic},
  \citenamefont {Bona-Casas}, \citenamefont {Bona}, \citenamefont {Rezzolla},\
  and\ \citenamefont {Palenzuela}}]{Alic:2011gg}%
  \BibitemOpen
  \bibfield  {author} {\bibinfo {author} {\bibfnamefont {D.}~\bibnamefont
  {Alic}}, \bibinfo {author} {\bibfnamefont {C.}~\bibnamefont {Bona-Casas}},
  \bibinfo {author} {\bibfnamefont {C.}~\bibnamefont {Bona}}, \bibinfo {author}
  {\bibfnamefont {L.}~\bibnamefont {Rezzolla}}, \ and\ \bibinfo {author}
  {\bibfnamefont {C.}~\bibnamefont {Palenzuela}},\ }\href {\doibase
  10.1103/PhysRevD.85.064040} {\bibfield  {journal} {\bibinfo  {journal} {Phys.
  Rev.}\ }\textbf {\bibinfo {volume} {D85}},\ \bibinfo {pages} {064040}
  (\bibinfo {year} {2012})},\ \Eprint {http://arxiv.org/abs/1106.2254}
  {arXiv:1106.2254 [gr-qc]} \BibitemShut {NoStop}%
\bibitem [{\citenamefont {L{\"o}ffler}\ \emph {et~al.}(2012)\citenamefont
  {L{\"o}ffler}, \citenamefont {Faber}, \citenamefont {Bentivegna},
  \citenamefont {Bode}, \citenamefont {Diener}, \citenamefont {Haas},
  \citenamefont {Hinder}, \citenamefont {Mundim}, \citenamefont {Ott},
  \citenamefont {Schnetter}, \citenamefont {Allen}, \citenamefont
  {Campanelli},\ and\ \citenamefont {Laguna}}]{Loffler:2011ay}%
  \BibitemOpen
  \bibfield  {author} {\bibinfo {author} {\bibfnamefont {F.}~\bibnamefont
  {L{\"o}ffler}}, \bibinfo {author} {\bibfnamefont {J.}~\bibnamefont {Faber}},
  \bibinfo {author} {\bibfnamefont {E.}~\bibnamefont {Bentivegna}}, \bibinfo
  {author} {\bibfnamefont {T.}~\bibnamefont {Bode}}, \bibinfo {author}
  {\bibfnamefont {P.}~\bibnamefont {Diener}}, \bibinfo {author} {\bibfnamefont
  {R.}~\bibnamefont {Haas}}, \bibinfo {author} {\bibfnamefont {I.}~\bibnamefont
  {Hinder}}, \bibinfo {author} {\bibfnamefont {B.~C.}\ \bibnamefont {Mundim}},
  \bibinfo {author} {\bibfnamefont {C.~D.}\ \bibnamefont {Ott}}, \bibinfo
  {author} {\bibfnamefont {E.}~\bibnamefont {Schnetter}}, \bibinfo {author}
  {\bibfnamefont {G.}~\bibnamefont {Allen}}, \bibinfo {author} {\bibfnamefont
  {M.}~\bibnamefont {Campanelli}}, \ and\ \bibinfo {author} {\bibfnamefont
  {P.}~\bibnamefont {Laguna}},\ }\href@noop {} {\bibfield  {journal} {\bibinfo
  {journal} {Class. Quant. Grav.}\ }\textbf {\bibinfo {volume} {29}},\ \bibinfo
  {pages} {115001} (\bibinfo {year} {2012})},\ \Eprint
  {http://arxiv.org/abs/1111.3344} {arXiv:1111.3344 [gr-qc]} \BibitemShut
  {NoStop}%
\bibitem [{ein()}]{einsteintoolkit}%
  \BibitemOpen
  \href@noop {} {}\bibinfo {note} {Einstein Toolkit home page: {\tt
  http://einsteintoolkit.org}}\BibitemShut {NoStop}%
\bibitem [{cac()}]{cactus_web}%
  \BibitemOpen
  \href@noop {} {}\bibinfo {note} {Cactus Computational Toolkit home page: {\tt
  http://cactuscode.org}}\BibitemShut {NoStop}%
\bibitem [{\citenamefont {Schnetter}\ \emph {et~al.}(2004)\citenamefont
  {Schnetter}, \citenamefont {Hawley},\ and\ \citenamefont
  {Hawke}}]{Schnetter-etal-03b}%
  \BibitemOpen
  \bibfield  {author} {\bibinfo {author} {\bibfnamefont {E.}~\bibnamefont
  {Schnetter}}, \bibinfo {author} {\bibfnamefont {S.~H.}\ \bibnamefont
  {Hawley}}, \ and\ \bibinfo {author} {\bibfnamefont {I.}~\bibnamefont
  {Hawke}},\ }\href@noop {} {\bibfield  {journal} {\bibinfo  {journal} {Class.
  Quant. Grav.}\ }\textbf {\bibinfo {volume} {21}},\ \bibinfo {pages} {1465}
  (\bibinfo {year} {2004})},\ \Eprint {http://arxiv.org/abs/gr-qc/0310042}
  {gr-qc/0310042} \BibitemShut {NoStop}%
\bibitem [{\citenamefont {Thornburg}(2004)}]{Thornburg2003:AH-finding}%
  \BibitemOpen
  \bibfield  {author} {\bibinfo {author} {\bibfnamefont {J.}~\bibnamefont
  {Thornburg}},\ }\href {\doibase 10.1088/0264-9381/21/2/026} {\bibfield
  {journal} {\bibinfo  {journal} {Class. Quant. Grav.}\ }\textbf {\bibinfo
  {volume} {21}},\ \bibinfo {pages} {743} (\bibinfo {year} {2004})},\ \Eprint
  {http://arxiv.org/abs/gr-qc/0306056} {gr-qc/0306056} \BibitemShut {NoStop}%
\bibitem [{\citenamefont {Dreyer}\ \emph {et~al.}(2003)\citenamefont {Dreyer},
  \citenamefont {Krishnan}, \citenamefont {Shoemaker},\ and\ \citenamefont
  {Schnetter}}]{Dreyer02a}%
  \BibitemOpen
  \bibfield  {author} {\bibinfo {author} {\bibfnamefont {O.}~\bibnamefont
  {Dreyer}}, \bibinfo {author} {\bibfnamefont {B.}~\bibnamefont {Krishnan}},
  \bibinfo {author} {\bibfnamefont {D.}~\bibnamefont {Shoemaker}}, \ and\
  \bibinfo {author} {\bibfnamefont {E.}~\bibnamefont {Schnetter}},\ }\href@noop
  {} {\bibfield  {journal} {\bibinfo  {journal} {Phys. Rev.}\ }\textbf
  {\bibinfo {volume} {D67}},\ \bibinfo {pages} {024018} (\bibinfo {year}
  {2003})},\ \Eprint {http://arxiv.org/abs/gr-qc/0206008} {gr-qc/0206008}
  \BibitemShut {NoStop}%
\bibitem [{\citenamefont {Campanelli}\ \emph
  {et~al.}(2007{\natexlab{c}})\citenamefont {Campanelli}, \citenamefont
  {Lousto}, \citenamefont {Zlochower}, \citenamefont {Krishnan},\ and\
  \citenamefont {Merritt}}]{Campanelli:2006fy}%
  \BibitemOpen
  \bibfield  {author} {\bibinfo {author} {\bibfnamefont {M.}~\bibnamefont
  {Campanelli}}, \bibinfo {author} {\bibfnamefont {C.~O.}\ \bibnamefont
  {Lousto}}, \bibinfo {author} {\bibfnamefont {Y.}~\bibnamefont {Zlochower}},
  \bibinfo {author} {\bibfnamefont {B.}~\bibnamefont {Krishnan}}, \ and\
  \bibinfo {author} {\bibfnamefont {D.}~\bibnamefont {Merritt}},\ }\href@noop
  {} {\bibfield  {journal} {\bibinfo  {journal} {Phys. Rev.}\ }\textbf
  {\bibinfo {volume} {D75}},\ \bibinfo {pages} {064030} (\bibinfo {year}
  {2007}{\natexlab{c}})},\ \Eprint {http://arxiv.org/abs/gr-qc/0612076}
  {gr-qc/0612076} \BibitemShut {NoStop}%
\bibitem [{\citenamefont {Campanelli}\ and\ \citenamefont
  {Lousto}(1999)}]{Campanelli:1998jv}%
  \BibitemOpen
  \bibfield  {author} {\bibinfo {author} {\bibfnamefont {M.}~\bibnamefont
  {Campanelli}}\ and\ \bibinfo {author} {\bibfnamefont {C.~O.}\ \bibnamefont
  {Lousto}},\ }\href {\doibase 10.1103/PhysRevD.59.124022} {\bibfield
  {journal} {\bibinfo  {journal} {Phys. Rev.}\ }\textbf {\bibinfo {volume}
  {D59}},\ \bibinfo {pages} {124022} (\bibinfo {year} {1999})},\ \Eprint
  {http://arxiv.org/abs/gr-qc/9811019} {arXiv:gr-qc/9811019 [gr-qc]}
  \BibitemShut {NoStop}%
\bibitem [{\citenamefont {Lousto}\ and\ \citenamefont
  {Zlochower}(2007)}]{Lousto:2007mh}%
  \BibitemOpen
  \bibfield  {author} {\bibinfo {author} {\bibfnamefont {C.~O.}\ \bibnamefont
  {Lousto}}\ and\ \bibinfo {author} {\bibfnamefont {Y.}~\bibnamefont
  {Zlochower}},\ }\href@noop {} {\bibfield  {journal} {\bibinfo  {journal}
  {Phys. Rev.}\ }\textbf {\bibinfo {volume} {D76}},\ \bibinfo {pages}
  {041502(R)} (\bibinfo {year} {2007})},\ \Eprint
  {http://arxiv.org/abs/gr-qc/0703061} {gr-qc/0703061} \BibitemShut {NoStop}%
\bibitem [{\citenamefont {Nakano}\ \emph {et~al.}(2015)\citenamefont {Nakano},
  \citenamefont {Healy}, \citenamefont {Lousto},\ and\ \citenamefont
  {Zlochower}}]{Nakano:2015pta}%
  \BibitemOpen
  \bibfield  {author} {\bibinfo {author} {\bibfnamefont {H.}~\bibnamefont
  {Nakano}}, \bibinfo {author} {\bibfnamefont {J.}~\bibnamefont {Healy}},
  \bibinfo {author} {\bibfnamefont {C.~O.}\ \bibnamefont {Lousto}}, \ and\
  \bibinfo {author} {\bibfnamefont {Y.}~\bibnamefont {Zlochower}},\ }\href
  {\doibase 10.1103/PhysRevD.91.104022} {\bibfield  {journal} {\bibinfo
  {journal} {Phys. Rev.}\ }\textbf {\bibinfo {volume} {D91}},\ \bibinfo {pages}
  {104022} (\bibinfo {year} {2015})},\ \Eprint
  {http://arxiv.org/abs/1503.00718} {arXiv:1503.00718 [gr-qc]} \BibitemShut
  {NoStop}%
\bibitem [{\citenamefont {Healy}\ \emph
  {et~al.}(2017{\natexlab{b}})\citenamefont {Healy}, \citenamefont {Lousto},
  \citenamefont {Nakano},\ and\ \citenamefont {Zlochower}}]{Healy:2017zqj}%
  \BibitemOpen
  \bibfield  {author} {\bibinfo {author} {\bibfnamefont {J.}~\bibnamefont
  {Healy}}, \bibinfo {author} {\bibfnamefont {C.~O.}\ \bibnamefont {Lousto}},
  \bibinfo {author} {\bibfnamefont {H.}~\bibnamefont {Nakano}}, \ and\ \bibinfo
  {author} {\bibfnamefont {Y.}~\bibnamefont {Zlochower}},\ }\href {\doibase
  10.1088/1361-6382/aa7929} {\bibfield  {journal} {\bibinfo  {journal} {Class.
  Quant. Grav.}\ }\textbf {\bibinfo {volume} {34}},\ \bibinfo {pages} {145011}
  (\bibinfo {year} {2017}{\natexlab{b}})},\ \Eprint
  {http://arxiv.org/abs/1702.00872} {arXiv:1702.00872 [gr-qc]} \BibitemShut
  {NoStop}%
\bibitem [{\citenamefont {Lousto}\ and\ \citenamefont
  {Zlochower}(2014)}]{Lousto:2013wta}%
  \BibitemOpen
  \bibfield  {author} {\bibinfo {author} {\bibfnamefont {C.~O.}\ \bibnamefont
  {Lousto}}\ and\ \bibinfo {author} {\bibfnamefont {Y.}~\bibnamefont
  {Zlochower}},\ }\href {\doibase 10.1103/PhysRevD.89.104052} {\bibfield
  {journal} {\bibinfo  {journal} {Phys. Rev.}\ }\textbf {\bibinfo {volume}
  {D89}},\ \bibinfo {pages} {104052} (\bibinfo {year} {2014})},\ \Eprint
  {http://arxiv.org/abs/1312.5775} {arXiv:1312.5775 [gr-qc]} \BibitemShut
  {NoStop}%
\bibitem [{\citenamefont {Lousto}\ and\ \citenamefont
  {Zlochower}(2011{\natexlab{b}})}]{Lousto:2010xk}%
  \BibitemOpen
  \bibfield  {author} {\bibinfo {author} {\bibfnamefont {C.~O.}\ \bibnamefont
  {Lousto}}\ and\ \bibinfo {author} {\bibfnamefont {Y.}~\bibnamefont
  {Zlochower}},\ }\href {\doibase 10.1103/PhysRevD.83.024003} {\bibfield
  {journal} {\bibinfo  {journal} {Phys. Rev.}\ }\textbf {\bibinfo {volume}
  {D83}},\ \bibinfo {pages} {024003} (\bibinfo {year} {2011}{\natexlab{b}})},\
  \Eprint {http://arxiv.org/abs/1011.0593} {arXiv:1011.0593 [gr-qc]}
  \BibitemShut {NoStop}%
\bibitem [{\citenamefont {Lousto}\ and\ \citenamefont
  {Zlochower}(2013)}]{Lousto:2012gt}%
  \BibitemOpen
  \bibfield  {author} {\bibinfo {author} {\bibfnamefont {C.~O.}\ \bibnamefont
  {Lousto}}\ and\ \bibinfo {author} {\bibfnamefont {Y.}~\bibnamefont
  {Zlochower}},\ }\href {\doibase 10.1103/PhysRevD.87.084027} {\bibfield
  {journal} {\bibinfo  {journal} {Phys. Rev.}\ }\textbf {\bibinfo {volume}
  {D87}},\ \bibinfo {pages} {084027} (\bibinfo {year} {2013})},\ \Eprint
  {http://arxiv.org/abs/1211.7099} {arXiv:1211.7099 [gr-qc]} \BibitemShut
  {NoStop}%
\bibitem [{\citenamefont {Healy}\ and\ \citenamefont
  {Lousto}(2018)}]{Healy:2018swt}%
  \BibitemOpen
  \bibfield  {author} {\bibinfo {author} {\bibfnamefont {J.}~\bibnamefont
  {Healy}}\ and\ \bibinfo {author} {\bibfnamefont {C.~O.}\ \bibnamefont
  {Lousto}},\ }\href {\doibase 10.1103/PhysRevD.97.084002} {\bibfield
  {journal} {\bibinfo  {journal} {Phys. Rev.}\ }\textbf {\bibinfo {volume}
  {D97}},\ \bibinfo {pages} {084002} (\bibinfo {year} {2018})},\ \Eprint
  {http://arxiv.org/abs/1801.08162} {arXiv:1801.08162 [gr-qc]} \BibitemShut
  {NoStop}%
\bibitem [{\citenamefont {Ruiz}\ \emph {et~al.}(2008)\citenamefont {Ruiz},
  \citenamefont {Takahashi}, \citenamefont {Alcubierre},\ and\ \citenamefont
  {Nunez}}]{Ruiz:2007yx}%
  \BibitemOpen
  \bibfield  {author} {\bibinfo {author} {\bibfnamefont {M.}~\bibnamefont
  {Ruiz}}, \bibinfo {author} {\bibfnamefont {R.}~\bibnamefont {Takahashi}},
  \bibinfo {author} {\bibfnamefont {M.}~\bibnamefont {Alcubierre}}, \ and\
  \bibinfo {author} {\bibfnamefont {D.}~\bibnamefont {Nunez}},\ }\href
  {\doibase 10.1007/s10714-007-0570-8} {\bibfield  {journal} {\bibinfo
  {journal} {Gen. Rel. Grav.}\ }\textbf {\bibinfo {volume} {40}},\ \bibinfo
  {pages} {2467} (\bibinfo {year} {2008})},\ \Eprint
  {http://arxiv.org/abs/0707.4654} {arXiv:0707.4654 [gr-qc]} \BibitemShut
  {NoStop}%
\bibitem [{\citenamefont {Healy}\ and\ \citenamefont
  {Lousto}(2017)}]{Healy:2016lce}%
  \BibitemOpen
  \bibfield  {author} {\bibinfo {author} {\bibfnamefont {J.}~\bibnamefont
  {Healy}}\ and\ \bibinfo {author} {\bibfnamefont {C.~O.}\ \bibnamefont
  {Lousto}},\ }\href {\doibase 10.1103/PhysRevD.95.024037} {\bibfield
  {journal} {\bibinfo  {journal} {Phys. Rev.}\ }\textbf {\bibinfo {volume}
  {D95}},\ \bibinfo {pages} {024037} (\bibinfo {year} {2017})},\ \Eprint
  {http://arxiv.org/abs/1610.09713} {arXiv:1610.09713 [gr-qc]} \BibitemShut
  {NoStop}%
\bibitem [{\citenamefont {Abbott}\ \emph
  {et~al.}(2016{\natexlab{a}})\citenamefont {Abbott} \emph
  {et~al.}}]{Abbott:2016apu}%
  \BibitemOpen
  \bibfield  {author} {\bibinfo {author} {\bibfnamefont {B.~P.}\ \bibnamefont
  {Abbott}} \emph {et~al.} (\bibinfo {collaboration} {Virgo, LIGO
  Scientific}),\ }\href {\doibase 10.1103/PhysRevD.94.064035} {\bibfield
  {journal} {\bibinfo  {journal} {Phys. Rev.}\ }\textbf {\bibinfo {volume}
  {D94}},\ \bibinfo {pages} {064035} (\bibinfo {year} {2016}{\natexlab{a}})},\
  \Eprint {http://arxiv.org/abs/1606.01262} {arXiv:1606.01262 [gr-qc]}
  \BibitemShut {NoStop}%
\bibitem [{\citenamefont {{Lange}}\ \emph {et~al.}(2018)\citenamefont
  {{Lange}}, \citenamefont {{O'Shaughnessy}},\ and\ \citenamefont
  {{Rizzo}}}]{2018arXiv180510457L}%
  \BibitemOpen
  \bibfield  {author} {\bibinfo {author} {\bibfnamefont {J.}~\bibnamefont
  {{Lange}}}, \bibinfo {author} {\bibfnamefont {R.}~\bibnamefont
  {{O'Shaughnessy}}}, \ and\ \bibinfo {author} {\bibfnamefont {M.}~\bibnamefont
  {{Rizzo}}},\ }\href@noop {} {\bibfield  {journal} {\bibinfo  {journal} {ArXiv
  e-prints}\ } (\bibinfo {year} {2018})},\ \Eprint
  {http://arxiv.org/abs/1805.10457} {arXiv:1805.10457 [gr-qc]} \BibitemShut
  {NoStop}%
\bibitem [{\citenamefont {Abbott}\ \emph {et~al.}(2018)\citenamefont {Abbott}
  \emph {et~al.}}]{LIGOScientific:2018mvr}%
  \BibitemOpen
  \bibfield  {author} {\bibinfo {author} {\bibfnamefont {B.~P.}\ \bibnamefont
  {Abbott}} \emph {et~al.} (\bibinfo {collaboration} {LIGO Scientific,
  Virgo}),\ }\href@noop {} {\  (\bibinfo {year} {2018})},\ \Eprint
  {http://arxiv.org/abs/1811.12907} {arXiv:1811.12907 [astro-ph.HE]}
  \BibitemShut {NoStop}%
\bibitem [{\citenamefont {Healy}\ \emph
  {et~al.}(2018{\natexlab{b}})\citenamefont {Healy} \emph
  {et~al.}}]{Healy:2017abq}%
  \BibitemOpen
  \bibfield  {author} {\bibinfo {author} {\bibfnamefont {J.}~\bibnamefont
  {Healy}} \emph {et~al.},\ }\href {\doibase 10.1103/PhysRevD.97.064027}
  {\bibfield  {journal} {\bibinfo  {journal} {Phys. Rev.}\ }\textbf {\bibinfo
  {volume} {D97}},\ \bibinfo {pages} {064027} (\bibinfo {year}
  {2018}{\natexlab{b}})},\ \Eprint {http://arxiv.org/abs/1712.05836}
  {arXiv:1712.05836 [gr-qc]} \BibitemShut {NoStop}%
\bibitem [{\citenamefont {Schmidt}\ \emph {et~al.}(2015)\citenamefont
  {Schmidt}, \citenamefont {Ohme},\ and\ \citenamefont
  {Hannam}}]{Schmidt:2014iyl}%
  \BibitemOpen
  \bibfield  {author} {\bibinfo {author} {\bibfnamefont {P.}~\bibnamefont
  {Schmidt}}, \bibinfo {author} {\bibfnamefont {F.}~\bibnamefont {Ohme}}, \
  and\ \bibinfo {author} {\bibfnamefont {M.}~\bibnamefont {Hannam}},\ }\href
  {\doibase 10.1103/PhysRevD.91.024043} {\bibfield  {journal} {\bibinfo
  {journal} {Phys. Rev.}\ }\textbf {\bibinfo {volume} {D91}},\ \bibinfo {pages}
  {024043} (\bibinfo {year} {2015})},\ \Eprint {http://arxiv.org/abs/1408.1810}
  {arXiv:1408.1810 [gr-qc]} \BibitemShut {NoStop}%
\bibitem [{\citenamefont {Chamberlain}\ \emph {et~al.}(2019)\citenamefont
  {Chamberlain}, \citenamefont {Moore}, \citenamefont {Gerosa},\ and\
  \citenamefont {Yunes}}]{Chamberlain:2018snj}%
  \BibitemOpen
  \bibfield  {author} {\bibinfo {author} {\bibfnamefont {K.}~\bibnamefont
  {Chamberlain}}, \bibinfo {author} {\bibfnamefont {C.~J.}\ \bibnamefont
  {Moore}}, \bibinfo {author} {\bibfnamefont {D.}~\bibnamefont {Gerosa}}, \
  and\ \bibinfo {author} {\bibfnamefont {N.}~\bibnamefont {Yunes}},\ }\href
  {\doibase 10.1103/PhysRevD.99.024025} {\bibfield  {journal} {\bibinfo
  {journal} {Phys. Rev.}\ }\textbf {\bibinfo {volume} {D99}},\ \bibinfo {pages}
  {024025} (\bibinfo {year} {2019})},\ \Eprint
  {http://arxiv.org/abs/1809.04799} {arXiv:1809.04799 [gr-qc]} \BibitemShut
  {NoStop}%
\bibitem [{\citenamefont {Gerosa}\ \emph {et~al.}(2018)\citenamefont {Gerosa},
  \citenamefont {H\'ebert},\ and\ \citenamefont {Stein}}]{Gerosa:2018qay}%
  \BibitemOpen
  \bibfield  {author} {\bibinfo {author} {\bibfnamefont {D.}~\bibnamefont
  {Gerosa}}, \bibinfo {author} {\bibfnamefont {F.}~\bibnamefont {H\'ebert}}, \
  and\ \bibinfo {author} {\bibfnamefont {L.~C.}\ \bibnamefont {Stein}},\ }\href
  {\doibase 10.1103/PhysRevD.97.104049} {\bibfield  {journal} {\bibinfo
  {journal} {Phys. Rev.}\ }\textbf {\bibinfo {volume} {D97}},\ \bibinfo {pages}
  {104049} (\bibinfo {year} {2018})},\ \Eprint
  {http://arxiv.org/abs/1802.04276} {arXiv:1802.04276 [gr-qc]} \BibitemShut
  {NoStop}%
\bibitem [{\citenamefont {Abbott}\ \emph
  {et~al.}(2016{\natexlab{b}})\citenamefont {Abbott} \emph
  {et~al.}}]{Abbott:2016izl}%
  \BibitemOpen
  \bibfield  {author} {\bibinfo {author} {\bibfnamefont {B.~P.}\ \bibnamefont
  {Abbott}} \emph {et~al.} (\bibinfo {collaboration} {Virgo, LIGO
  Scientific}),\ }\href {\doibase 10.1103/PhysRevX.6.041014} {\bibfield
  {journal} {\bibinfo  {journal} {Phys. Rev.}\ }\textbf {\bibinfo {volume}
  {X6}},\ \bibinfo {pages} {041014} (\bibinfo {year} {2016}{\natexlab{b}})},\
  \Eprint {http://arxiv.org/abs/1606.01210} {arXiv:1606.01210 [gr-qc]}
  \BibitemShut {NoStop}%
\bibitem [{\citenamefont {Vallisneri}\ \emph {et~al.}(2015)\citenamefont
  {Vallisneri}, \citenamefont {Kanner}, \citenamefont {Williams}, \citenamefont
  {Weinstein},\ and\ \citenamefont {Stephens}}]{Vallisneri:2014vxa}%
  \BibitemOpen
  \bibfield  {author} {\bibinfo {author} {\bibfnamefont {M.}~\bibnamefont
  {Vallisneri}}, \bibinfo {author} {\bibfnamefont {J.}~\bibnamefont {Kanner}},
  \bibinfo {author} {\bibfnamefont {R.}~\bibnamefont {Williams}}, \bibinfo
  {author} {\bibfnamefont {A.}~\bibnamefont {Weinstein}}, \ and\ \bibinfo
  {author} {\bibfnamefont {B.}~\bibnamefont {Stephens}},\ }\bibfield
  {booktitle} {\emph {\bibinfo {booktitle} {{Proceedings, 10th International
  LISA Symposium: Gainesville, Florida, USA, May 18-23, 2014}}},\ }\href
  {\doibase 10.1088/1742-6596/610/1/012021} {\bibfield  {journal} {\bibinfo
  {journal} {J. Phys. Conf. Ser.}\ }\textbf {\bibinfo {volume} {610}},\
  \bibinfo {pages} {012021} (\bibinfo {year} {2015})},\ \Eprint
  {http://arxiv.org/abs/1410.4839} {arXiv:1410.4839 [gr-qc]} \BibitemShut
  {NoStop}%
\bibitem [{\citenamefont {{Pankow}}\ \emph {et~al.}(2015)\citenamefont
  {{Pankow}}, \citenamefont {{Brady}}, \citenamefont {{Ochsner}},\ and\
  \citenamefont {{O'Shaughnessy}}}]{2015PhRvD..92b3002P}%
  \BibitemOpen
  \bibfield  {author} {\bibinfo {author} {\bibfnamefont {C.}~\bibnamefont
  {{Pankow}}}, \bibinfo {author} {\bibfnamefont {P.}~\bibnamefont {{Brady}}},
  \bibinfo {author} {\bibfnamefont {E.}~\bibnamefont {{Ochsner}}}, \ and\
  \bibinfo {author} {\bibfnamefont {R.}~\bibnamefont {{O'Shaughnessy}}},\
  }\href {\doibase 10.1103/PhysRevD.92.023002} {\bibfield  {journal} {\bibinfo
  {journal} {\prd}\ }\textbf {\bibinfo {volume} {92}},\ \bibinfo {eid} {023002}
  (\bibinfo {year} {2015})},\ \Eprint {http://arxiv.org/abs/1502.04370}
  {arXiv:1502.04370 [gr-qc]} \BibitemShut {NoStop}%
\bibitem [{\citenamefont {Lange}\ \emph {et~al.}(2018)\citenamefont {Lange},
  \citenamefont {O'Shaughnessy},\ and\ \citenamefont {Rizzo}}]{Lange:2018pyp}%
  \BibitemOpen
  \bibfield  {author} {\bibinfo {author} {\bibfnamefont {J.}~\bibnamefont
  {Lange}}, \bibinfo {author} {\bibfnamefont {R.}~\bibnamefont
  {O'Shaughnessy}}, \ and\ \bibinfo {author} {\bibfnamefont {M.}~\bibnamefont
  {Rizzo}},\ }\href@noop {} {\  (\bibinfo {year} {2018})},\ \Eprint
  {http://arxiv.org/abs/1805.10457} {arXiv:1805.10457 [gr-qc]} \BibitemShut
  {NoStop}%
\end{thebibliography}%

\end{document}